\documentclass[12pt]{JHEP3}
\usepackage{amsfonts, amsmath, amsthm}
\usepackage{graphicx}

\title{Scanning the Parameter Space of   Holographic Superconductors}
\author{Obinna C. Umeh \\  Department of Mathematics and Applied Mathematics,\\ University of Cape Town,\\ Rondebosch, 7701,\\ Cape Town,  South Africa.\\umeobinna@gmail.com}

\abstract{We study various physical quantities associated with holographic s-wave superconductors as functions of  the scaling dimensions of the dual condensates. A bulk scalar field with negative mass squared $m^2$, satisfying the Breitenlohner-Freedman stability bound and the unitarity bound, and allowed to vary in $0.5$ unit intervals, were considered. We observe that all the  physical quantities investigated are sensitive to the scaling dimensions of the dual condensates. For all the $m^2$, the characteristic lengths diverge  at the critical temperature in agreement with the Ginzburg-Landau theory. The Ginzburg-Landau parameter, obtained from these length scales indicates that the  holographic superconductors can be type I or type II depending on the charge and the scaling dimensions of the dual condensates.  For a fixed charge, there exists a critical scaling dimension, above which a holographic  superconductor is type I, below which it becomes a type II.
}
 \keywords{Holographic superconductors, AdS/CFT correspondence} 
\preprint{CGG170709}

\begin{document}

\maketitle

\newpage
\section{Introduction}
The correspondence between gravitational theories in anti-de Sitter spacetime and certain quantum field theories~\cite{Maldacena:1997re} provides a unique way in which to study the strongly coupled sector of many quantum field theories. This remarkable result from string theory has allowed some insight~\cite{Kovtun:2004de} to be gained into why the quark-gluon plasma produced at the relativistic heavy ion collider (RHIC) behaves  like an almost  perfect fluid~\cite{Luzum:2008cw} (in contrast to the prediction of a high viscosity by perturbative quantum chromodynamics (QCD)~\cite{Baym:1990uj}). 
This remarkable result inspired the application of AdS/CFT techniques to certain condensed matter systems. Phenomena  such as the Hall effect and the Nernst effect  appear to have their dual gravitational descriptions~\cite{Hartnoll:2007ip,Hartnoll:2007ih,Hartnoll:2007ai}.

This technique has been employed recently, to shed some light on strongly coupled systems that undergo superconducting instabilities at a critical temperature (see~\cite{Hartnoll:2009sz,Herzog:2009xv} for a review).
It is understood~\cite{Sachdev:1999,Hertz:1976zz} that a quantum field theoretic description of condensed matter systems is possible in the vicinity of the quantum critical point (QCP), where the relevant scale invariant theories are similar to field theories describing second-order phase transitions, for example Ginzburg-Landau theory. As the QCP is approached, systems\footnote{For example, spin systems.} with the dynamical critical exponent $z=1$ become invariant under re-scalings of time and distance. This scale invariant symmetry forms part of the larger conformal symmetry group $SO(d+1,2)$~\cite{Sachdev:2009ve,Herzog:2009xv} of the quantum field theory, where $d$ is the number of spatial dimensions. The emergence of this symmetry near the QCP implies that its dual gravitational description must reside in anti-de Sitter spacetime with an  additional spatial dimension~\cite{Hartnoll:2009sz,Sachdev:2008ba}. 

According to the model of holographic superconductivity proposed in~\cite{Hartnoll:2008vx}, one can study strongly coupled s-wave superconductors, at a finite temperature and chemical potential, by considering a gravitational theory with an action  which has a black hole solution. The black hole, in this case, is charged under a $U(1)$ gauge field with a minimally coupled  complex scalar field $\Psi$. The no hair theorem does not apply if the scalar field has a non-trivial coupling to the gauge field~\cite{Gubser:2005ih}.
In this set up, the symmetry breaking in the bulk theory, which corresponds to a quantum phase transition to the superconducting phase in the boundary theory, is triggered by a position dependent negative mass squared formed from the gauge covariant derivative~\cite{Gubser:2008px}. Its contribution becomes significant near the horizon of the black hole, thereby forcing the scalar field to condense.

This model has been studied in various limits by several authors. For example, the authors of~\cite{Albash:2008eh,Albash:2009ix,Albash:2009iq,Montull:2009fe,Nakano:2008xc,Wen:2008pb} mapped the phase diagram of the holographic superconductors in the presence of an external magnetic field. They  also found and analyzed the physical properties of the vortex and droplets solutions for a scalar field with $m^2 l^2=-2$ ($l$ will be defined defined shortly.). The hydrodynamics of holographic superconductors was studied in detail in~\cite{Amado:2009ts}.  The effect of a vector current on the order of the phase transition was explored in~\cite{Basu:2008st}. The authors of~\cite{Horowitz:2008bn} showed that superconductivity is possible for a scalar field of various masses in $d=3$ and $d=4$ bulk dimensions. A proposal on how to calculate the superconducting characteristic length analytically, in the vicinity of QCP, was suggested in~\cite{Maeda:2008ir}. The effects of gravitational backreaction were considered, and a study made, for $m^2l^2=-2$, of the type of the holographic superconductors in~\cite{Gubser:2008pf,Hartnoll:2008kx}. So far there has not been any work which discusses  the relationship between the physical quantities associated with  the model and the scaling dimensions of the dual condensates.

The objective here is to go beyond the extension of the model already discussed in~\cite{Horowitz:2008bn} and to include a wider range of values of $ m^2$, satisfying the Brietenlohner-Freedman (BF) stability bound~\cite{Breitenlohner:1982jf} and the unitarity bound. We find it most convenient to choose values of $m^2$ in the interval of $0.5$ units. We shall focus our attention primarily on scalar fields with fall-offs at the AdS boundary, which are normalizable. Based on this behavior at the boundary, the scalar field $\Psi$ naturally split into two pieces, $\Psi_{\lambda_{-}}$ and $\Psi_{\lambda_{+}}$, with slower and faster fall-offs respectively. These describes different condensates with distinct superconducting phases and different scaling dimensions. We shall calculate each physical quantity associated with the condensates at a fixed temperature and for each value of $m^2$, which will allow us to ascertain the dependence of this physical quantity on the scaling dimension.

This report is organized as follows: In section \ref{chap3section2}, we define our conventions and derive the equations of motion. In section \ref{chap3section3}, we show that the superconducting phase of holographic superconductors of the class $\Psi_{\lambda_{-}}$ is very different from that of the class $\Psi_{\lambda_{+}}$. We present a discussion of the conductivity in section \ref{chap3section4} and show that in the limit in which the frequency $\omega$  approaches zero ($\omega\approx 0$), the superfluid density can be obtained from the frequency dependent conductivity. In  section \ref{chap3section5}, we solve the equations of motion perturbatively in order to calculate the characteristic lengths and the Ginzburg-Landau parameter. The conclusion is provided in section \ref{conclude}, while various results relating to the conductivity in the  boundary theory are presented in the appendices.

\section{Background Equations of Motion}\label{chap3section2}
The action of a gravitational theory with a $d+1$ black hole solution in anti de Sitter spacetime $AdS_{d+1}$
coupled to a matter field is given by
\begin{equation}\label{eqb1a}
I=I_{EH}+I_{matter},
\end{equation}
where $I_{EH}$ is the Einstein-Hilbert action with a negative cosmological constant $\Lambda$
\begin{equation}
I_{EH}=\frac{1}{2\kappa^{2}_{d}}\int d^{d+1}x\sqrt{-g}\left\lbrace R + \frac{d(d-1)}{2l^{2}}\right\rbrace,
\end{equation}
with $\kappa_{d}$  related to Newton's gravitational constant in $d-$dimensions $\kappa_{d}=8\pi G_{N}$. The cosmological constant $\Lambda$ depends on the radius of curvature  of the anti de Sitter spacetime, $l$,  $\Lambda=d(d-1)/2l^{2}$.
$I_{matter}$ is the action for the Abelian Higgs system expanded to quadratic order in the scalar field
\begin{equation}\label{eqb1}
I_{matter}=\frac{1}{2\kappa^{2}_{d}}\int d^{d+1}x\sqrt{-g}\left\lbrace   -\frac{1}{4} F^{\mu \nu} F_{\mu \nu} - |\partial \Psi-iqA\Psi|^{2}-m^{2}|\Psi |^{2} \right\rbrace ,
\end{equation}
where the gauge field and the scalar field are coupled  through the gauge covariant derivative, $ D_{\mu}= \partial _{\mu} + iqA_{\mu}$. Here $\partial _{\mu}$ is the spacetime covariant derivative, $A_{\mu}$ is the gauge field, with associated field strength $F_{\mu\nu}$, and $\Psi$ is a complex scalar field.
In the probe limit, the matter field can be  re-scaled as 
\begin{eqnarray}
A_{\mu}\rightarrow A_{\mu}/q \\ \nonumber
\Psi \rightarrow \Psi/q,
\end{eqnarray}
which ensures that the quadratic potential scales as $V\left( |\Psi|^2\right) \rightarrow V\left( |\Psi|^2\right)/q^{2}$ and the entire matter action as $I_{matter}\rightarrow I_{matter}/q^{2}$. 
In the limit $q \rightarrow \infty$, the action for Abelian-Higgs system $I_{matter}$ decouples from the Einstein-Hilbert action $I_{EH}$.
As noted in~\cite{Hartnoll:2008vx}, the probe approximation remains  valid as long as $\Psi$ and scalar potential $\Phi$ are not  large in the Planck limit.  Another way to implement the probe approximation suggested in~\cite{Yarom:2009uq}, is to consider a formal expansion of the full backreacted geometry in inverse powers of $q$. Then the leading order matter solutions will depend on $q$ as $\mathcal{O}(q^{-1})$, while the leading order metric $\mathcal{O}(q^{2})$  receives $\mathcal{O}(q^{-2})$  corrections.

The equations of motion for the  scalar field and Maxwell fields reads
\begin{equation}\label{eqb3}
\frac{1}{\sqrt{-g}}D_{\mu}\left( \sqrt{-g}g^{\mu\nu} D _{\nu} \Psi \right) = m^{2}\Psi,
\end{equation}
\begin{equation}\label{eqb4}
\frac{1}{\sqrt{-g}}\partial_{\mu}\left( \sqrt{-g}g^{\nu\lambda}g^{\mu\sigma}F_{\lambda\sigma}\right)  =g^{\mu\nu}J_{\mu},
\end{equation}
where the current $J_{\mu}$ is given by
\begin{equation}
 J_{\mu}=\left( i\left( \Psi \bar \partial _{\mu}\Psi-\partial _{\mu}\Psi \bar \Psi\right)  + 2A_{\mu}\Psi \bar \Psi\right).
\end{equation}
We consider the  $d+1$ planar black hole ansatz
\begin{equation}\label{eqb5}
ds^{2}= -f(r)dt^{2}+ \frac{dr^{2}}{f(r)} + r^{2}dx_{i}dx^{i},
\end{equation}
 where $ f(r) = \frac{r^{2}}{l^{2}}(1-\frac{r_{0}^{d}}{r^{d}})$ and $i$ runs from $1$ to $(d-2)$. Here $r=r_{0}$ is the event horizon and the Hawking temperature of the black hole  is given by
\begin{equation}\label{eqb6a}
T= \frac{ r_{0}d}{4\pi l^{2}}.
\end{equation}
It is more convenient to make a change of coordinates $ z=r_{0}/r$,  so that the  metric (\ref{eqb5}) becomes
\begin{equation}\label{eqb7}
ds^{2} = \frac{l^{2}\alpha(T)}{z^{2}}\left( -h(z)dt^{2} + dx_{i}dx^{i}\right)  + \frac{l^{2}dz^{2}}{z^{2}h(z)},
\end{equation}
where $\alpha(T) \equiv 4\pi T=r_{0} d/l^{2}$ and $h(z)= (1-z^{d})$. Here  $z=1$ and  $z=0$  is the  event horizon and AdS boundary  respectively.
We consider the following  ansatze\footnote{From these ansatze  we can see that the  phase of the scalar field is fixed.} for the matter fields $A_{\mu}dx^{\mu}=\Phi(z) dt$  and $\Psi = \Psi(z)$.
Using the ansatze in  the  equations of motion, the scalar and gauge fields yield respectively
\begin{equation}\label{eqb8}
\Psi''+\left( \frac{h'}{h}+ \frac{d-1}{z}\right) \Psi' + \frac{\tilde \Phi^{2} \Psi}{h^{2}} -\frac{m^{2}}{hz^{2}}\Psi =0,
\end{equation}
and
\begin{equation}\label{eqb9}
\tilde \Phi''- \frac{d-3}{z}\tilde \Phi' -\frac{2\Psi^{2}}{hz^{2}}\tilde \Phi = 0,
\end{equation}
where $\tilde \Phi\equiv \Phi / \alpha(T) $ and $l=1$. 
Regularity  at the horizon requires
\begin{eqnarray}
\Psi'\big|_{z=1}=\frac{m^{2}\Psi}{d}\big|_{z=1}, \\ \nonumber
\tilde \Phi\big|_{z=1} =0.
\end{eqnarray}
Near the $AdS$ boundary  the  scalar field and  the scalar potential behave as 
\begin{eqnarray}\label{eqb16}
\Psi = \Psi_{\lambda_{-}} z^{\lambda_{-}}+\Psi_{\lambda_{+}}z^{\lambda_{+}}+... \\ \nonumber
\tilde \Phi = \mu -\rho z^{d-2}+... ,
\end{eqnarray}
where $\lambda$  is the  dimension of the dual operator, which satisfies  the relation
\begin{equation}\label{eqb17}
\lambda\left( \lambda-d\right) =m^{2},
\end{equation} 
with solutions 
$\lambda_{\pm}= \frac{1}{2}\left( d \pm \sqrt{d^{2}+4m^{2}}\right)$. The stability of AdS vacuum, requires that the scalar field of negative mass squared must satisfy the BF bound~\cite {Breitenlohner:1982jf}, $m^{2} \geq -d^{2}/4,$ and in general the unitarity bound~\cite{Klebanov:1999tb}, $\lambda\geq(d-2)/2$. In the analysis that follows, we consider the values of $m^2$ within the range $-d^2/4\leq m^2 < -d^2/4+1$. Both modes of the asymptotic values of the scalar fields whose $m^2$ are within this range are normalizable, except at the saturation of the BF bound. For $m^2\geq d^2/4 +1$, only the $\lambda_{+}$ is normalizable, since $\lambda_{-}$ is below the unitarity bound. As mentioned in the introduction our primary  focus is on the scalar fields with $m^2$ within this range $\left( -d^2/4\leq m^2 < -d^2/4+1\right)$, which we can achieve by considering  $m^2$ in $0.5$ unit interval. The fixed interval makes the analysis and interpretation of the results less challenging.

The AdS/CFT dictionary~\cite{Witten:1998qj,Gubser:1998bc}  relates the constant coefficients of the asymptotic solutions (equation (\ref{eqb16}))  to physical quantities in the boundary theory. The coefficients $\Psi_{\lambda i}$ are coefficients of the normalizable modes of the scalar field equation, they both correspond to expectation values in the dual field theory $\Psi_{\lambda i}=\left\langle \mathcal{O}_{\lambda i}\right\rangle $.  $\mu$ and $\rho $  correspond to the chemical potential and charge density in the dual field theory, respectively.

\section{Phase Transitions for Various Condensates}\label{chap3section3}
\begin{figure}[htb!]
\includegraphics[scale=0.65]{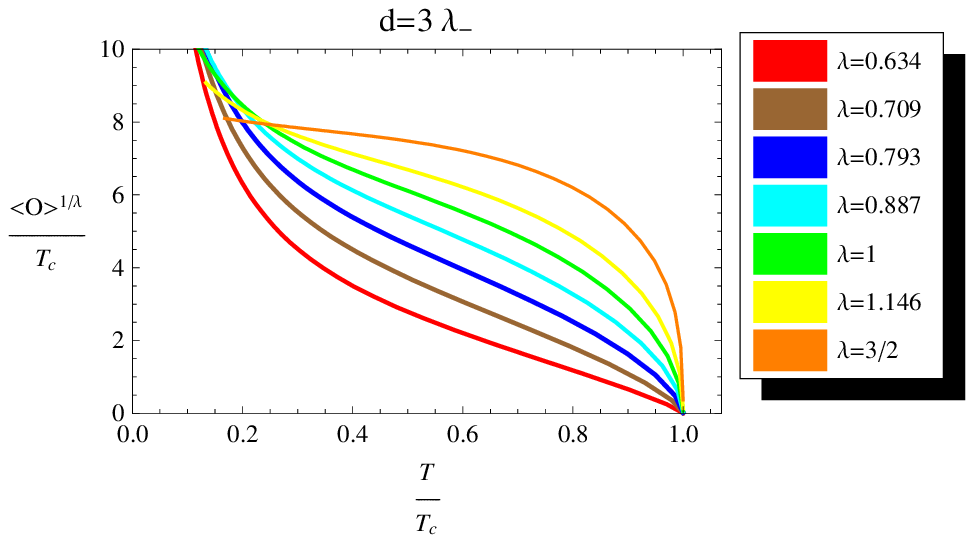}
\includegraphics[scale=0.65]{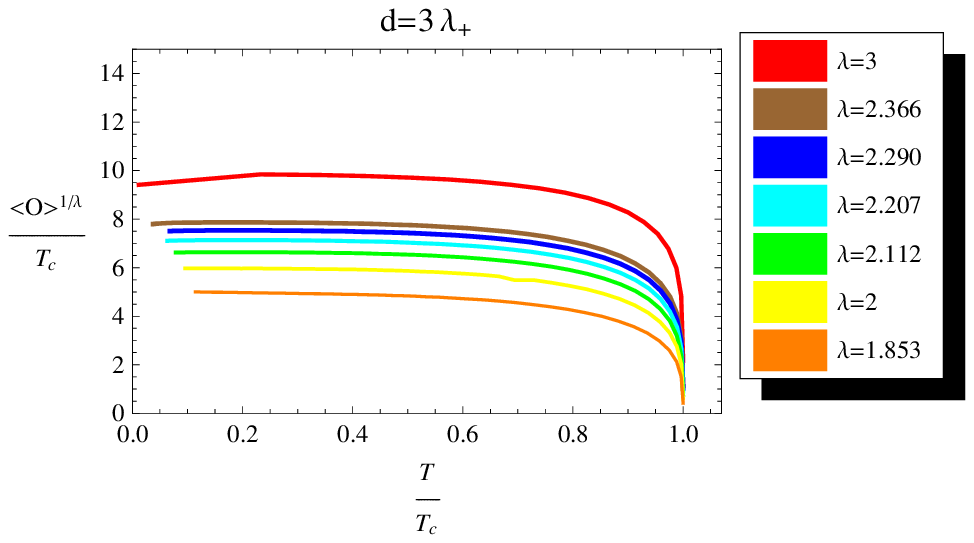}
\begin{center}
\includegraphics[scale=0.65]{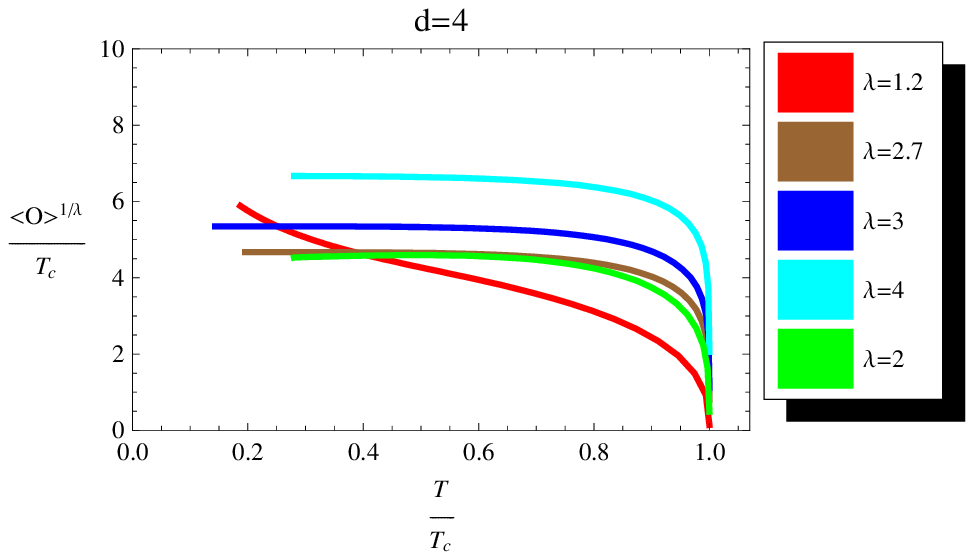}
\caption{The condensates as a function of temperature for various condensates $ \mathcal{O}_{\lambda} $ in the boundary theory. The figures are labelled by the scaling dimensions of the dual condensates. The  upper left graphs are condensates dual to the modes of the scalar field with slower fall-off $\Psi_{\lambda_{-}}$, while the upper right figure  shows the condensates dual to the modes of scalar fields with faster fall-off $\Psi_{\lambda_{+}}$. The graphs are labelled by $\lambda_{-}$ and $\lambda_{+}$ to distinguish between the two classes of condensates in the $2+1$ boundary theory. Below the two graphs is the condensates as a function of temperature for various condensate $\mathcal{O}_{\lambda}$ in $3+1$ dual field theory }\label{chap3Figure3}
\end{center}
\end{figure}

Apart from the trivial solutions $\Psi=0$ and $\tilde \Phi= \mu -\rho z^{d-2}$,
a non-trivial solution to equations (\ref{eqb8}) and (\ref{eqb9}) which describe the superconducting phase in the dual field theory, exist below a critical temperature.  The critical temperature is defined, for $\Psi_{\lambda_{-}}$, when $\Psi_{\lambda_{+}}$ vanishes and for  $\Psi_{\lambda_{+}}$, when $\Psi_{\lambda_{-}}$ vanishes. 
We present the  solutions to equations (\ref{eqb8}) and (\ref{eqb9}) obtained numerically in figure \ref{chap3Figure3}.

The temperature scales as $T\sim \rho^{1/2}$ and $T\sim \rho^{1/3}$ in the $2+1$ and $3+1$ boundary theory respectively.
Notice that the  condensates of the class $\Psi_{\lambda_{-}}$ converge at $\left\langle \mathcal{O}\right\rangle /T_c\approx 10$ before they  collectively diverge. The signatures of the divergence near zero temperature become more pronounced as  $\lambda$  approaches  the unitarity bound. A similar divergence was observed in~\cite{Hartnoll:2008vx} for $\lambda=1$ and  was attributed to the  probe approximation. But recent study~\cite{Hartnoll:2008kx} which considered gravitational backreaction, also show some signatures of  divergence for $\lambda=1$ when the charge $q$ becomes large.  This divergence might be an artifact of large N. There are obvious differences between the superconducting phase of $\Psi_{\lambda_{-}}$ and that of $\Psi_{\lambda_{+}}$. The condensates of the class $\Psi_{\lambda_{-}}$ show a gradual transition to the superconducting phase\footnote{In $d=4$ bulk dimensions the range of permissible values of $m^2$ is small hence we did not distinguish between the two classes in the graphical representation.  All the features as explained for the  $2+1$ boundary theory are also present.}.

The amount of condensate in each case can be calculated from the numerical solutions to equations (\ref{eqb8}) and (\ref{eqb9}) at a fixed temperature $T/T_c$, in the vicinity of QCP. The results are shown in figure \ref{chap3Figure3} (right).
\begin{figure}[htb!]
\includegraphics[scale=0.65]{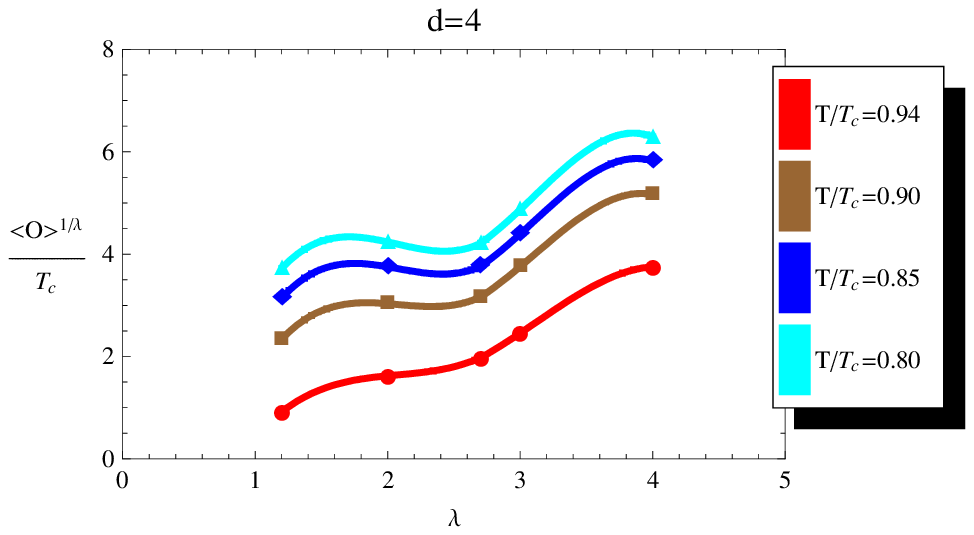}
\includegraphics[scale=0.65]{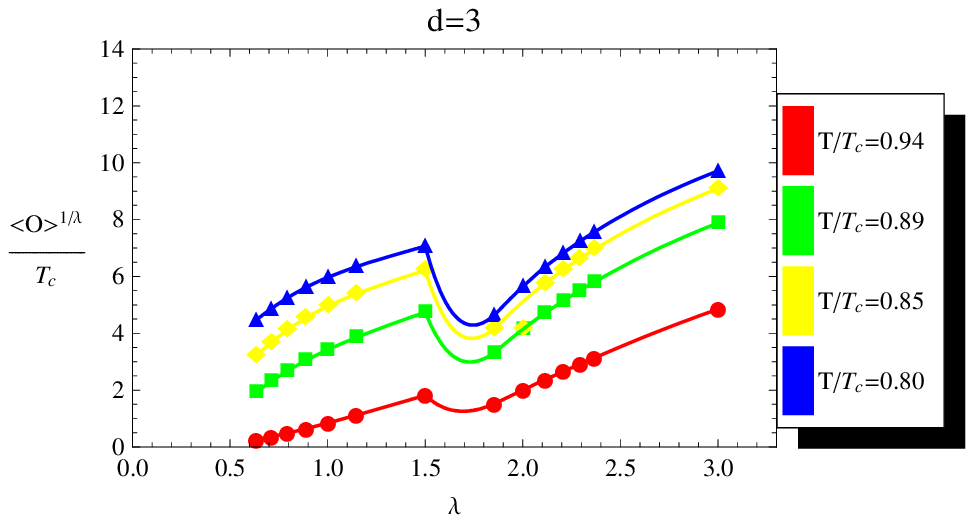}
\caption{The amount of condensates as a function of the dimension of the dual operator (right) computed at different fixed temperatures. The dots represent the actual value and the continuous line is an interpolation between the actual values. This from of representation is used in the rest of the report.} 
\label{chap3Figure4}
\end{figure}
There appears to be a discontinuity in the amount of condensates between holographic superconductors of the class $\Psi_{\lambda_{-}}$   and  that of  class $\Psi_{\lambda_{+}}$  at $\lambda_{crit} = \lambda_{BF}$ in both $2+1$ and $3+1$ boundary theories. This might be an indication that the two classes have different superconducting coherence factors~\cite{Hartnoll:2008vx}. The height of the discontinuous gap increases as the temperature decreases.

\begin{figure}[htb!]
\includegraphics[scale=0.65]{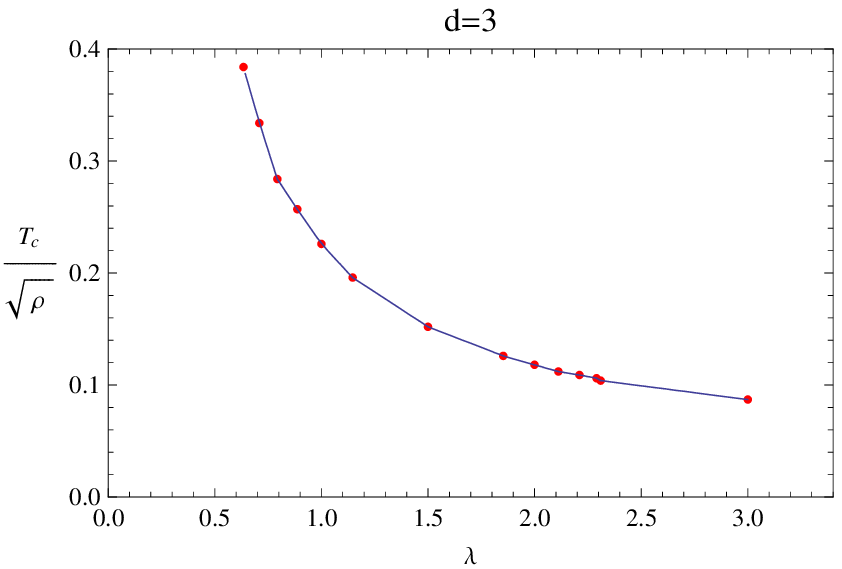}
\includegraphics[scale=0.65]{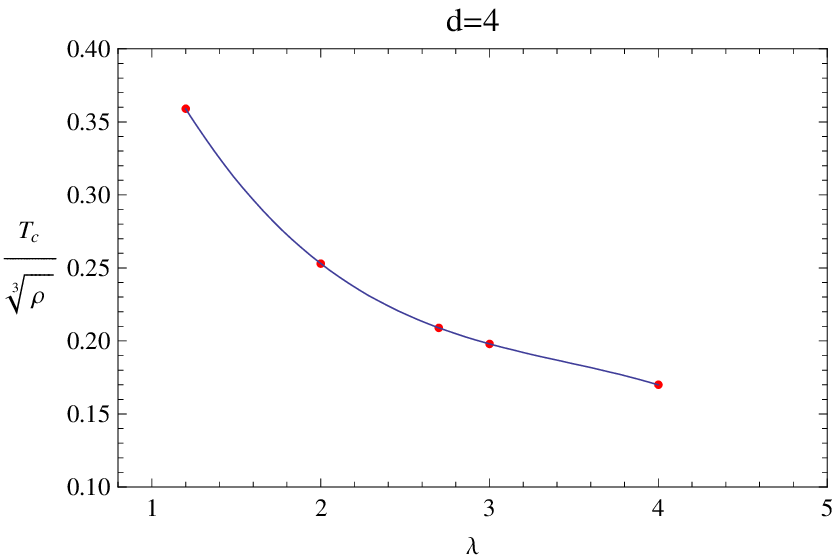}
\caption{The dependence of the critical temperature on the dimensions of the dual condensates in $2+1$ and $3+1$ boundary theories}. 
\label{chap3Figure4a}
\end{figure}

 The dependence of the critical temperatures for various condensates on the dimension of the dual operator is shown in figure  \ref{chap3Figure4a}.
The condensates with high scaling dimensions have  relatively very low critical temperature. In general as the critical  temperature decreases as the dimension of dual condensate increases in both the $3+1$ and $2+1$ boundary theories.

\section{Conductivity}\label{chap3section4}
Within the frame work of the AdS/CFT correspondence, the conductivity in the boundary theory can be calculated from the  Maxwell field in the bulk theory. This can be done in the probe limit by perturbing the Maxwell field at zero spatial momentum on the fixed black hole background: 
With  the ansatz for the perturbed Maxwell field,
 $\delta  A_{x}=A_{x}(z) e^{i\omega t} dx$,
a linearized equation of motion  results
\begin{equation}\label{eqb21}
 A_{x}''+\left( \frac{h'}{h}-\frac{d-3}{z}\right)A_{x}' +\left( \frac{\omega}{h^{2}}-\frac{2\Psi^{2}}{z^{2}h}\right) A_{x}=0.
\end{equation} 
Equation (\ref{eqb21}) is  solved with an ingoing wave boundary condition~\cite {Son:2002sd}  near the  horizon of the black hole in order to suppress near horizon oscillations:
\begin{equation}\label{eqb22}
 A_{x}(z)=h(z)^{-4\pi i\omega /T} A_{x}(z).
\end{equation}

\subsection{Conductivity in the ($2+1$)-dimensional dual field theory}
In  an odd number of dimensions (e.g. $d=3$)  the solution to the Maxwell's equation (\ref{eqb21}) behaves  near the boundary as 
\begin{equation}
 A_{x}= A^{(0)}+ A^{(1)} z+ ...
\end{equation}
From Ohm's law and  the dictionary of  AdS/CFT correspondence, the conductivity becomes
\begin{equation}
 \sigma(\omega)=\frac{A^{(1)}}{i\omega A^{(0)}}.
\end{equation}
 The  plots of the real and imaginary part of the conductivity  against the frequency normalized by individual  condensate are shown in  appendix \ref{appA},  figure \ref{Figure4c}  and appendix \ref{appC}, figure \ref{Figure4d} for the two classes of holographic superconductors in the $2+1$ boundary theory. 

\subsection{Conductivity in the ($3+1$)-dimensional dual field theory}
When the bulk  dimension  is even (e.g. $d=4$), there exists a logarithmic divergence of the Maxwell's field in the action \ref{eqb1a}: 
\begin{equation}
 A_{x}=A^{(0)} + A^{(2)}z^{2} + A^{(0)}\omega^{2}z^{2}\log \frac{\Lambda}{z}.
\end{equation}
A boundary counter term may be  added to remove the divergence~\cite{Taylor:2000xw},
so that  the conductivity becomes~\cite{Horowitz:2008bn}
\begin{equation}
 \sigma(\omega)=\frac{2A^{(2)}}{i\omega A^{(0)}}+\frac{i\omega}{2}
\end{equation}
The numerical solutions to equation (\ref{eqb21}) in the $3+1$ boundary theory, is shown 
 in appendix \ref{appB}, figure \ref{Figure6} and  for the  frequency  normalized by the individual superconducting condensate.
We could not resolve the delta function at $\omega=0$  numerically. However, it can be seen from the  Kramers-Kronig relation
\begin{equation}
 Im[\sigma(\omega)]= -\frac{1}{\pi}\mathcal{P}\int_{-\infty}^{\infty}\frac{Re[\sigma(\omega')]d\omega'}{\omega'-\omega},
\end{equation}
that there is a delta function  at  $\omega=0$ for all the condensates, since $\omega=0$ is a pole in the imaginary part of the conductivity.
The gap frequency $\omega_{g}$ remain approximately the same for all the condensates $\omega_{g}/T_{c} \approx 8$, irrespective of the number of bulk dimensions.

\subsection{Superfluid density and magnetic penetration depth}
\begin{figure}[htb!]
\includegraphics[scale=0.65]{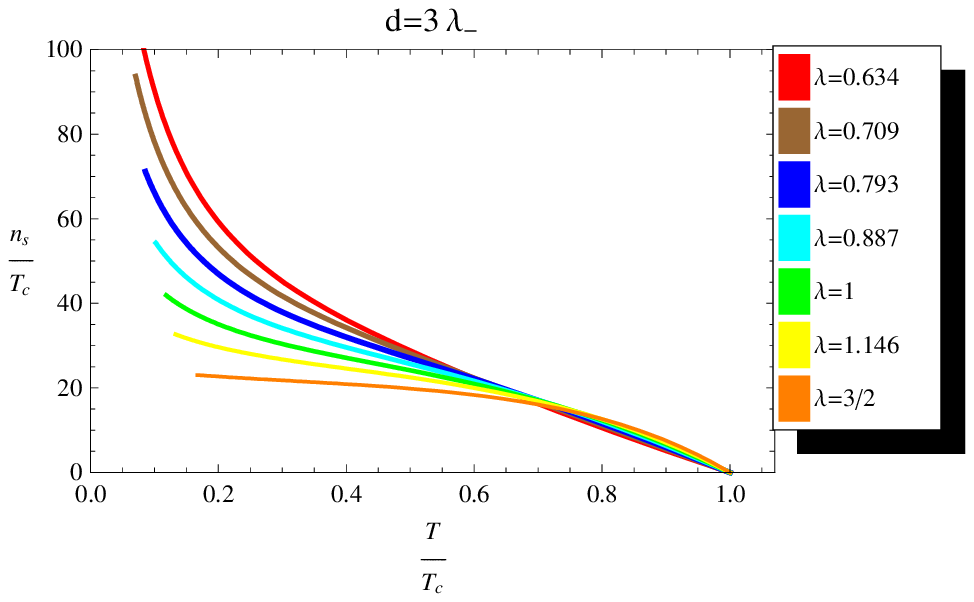}
\includegraphics[scale=0.65]{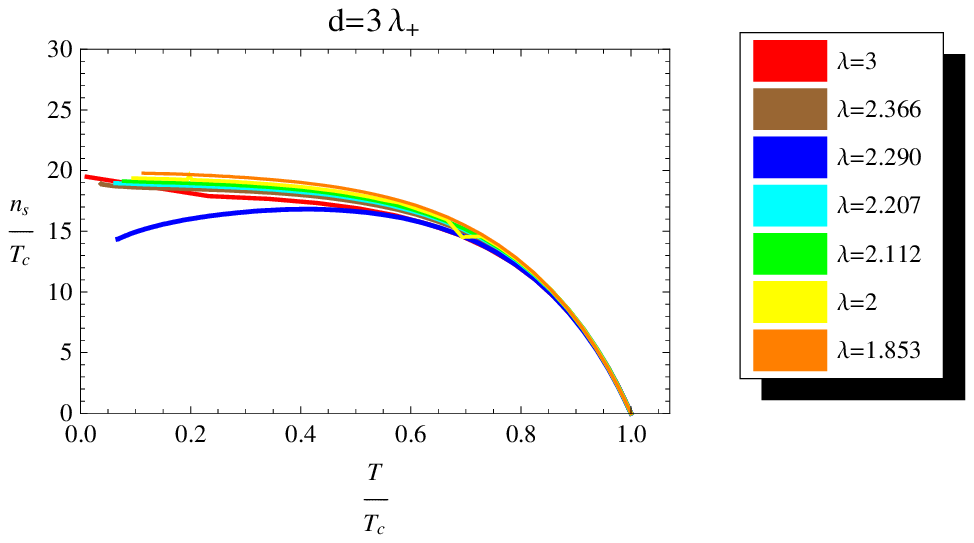}
\begin{center}
\includegraphics[scale=0.65]{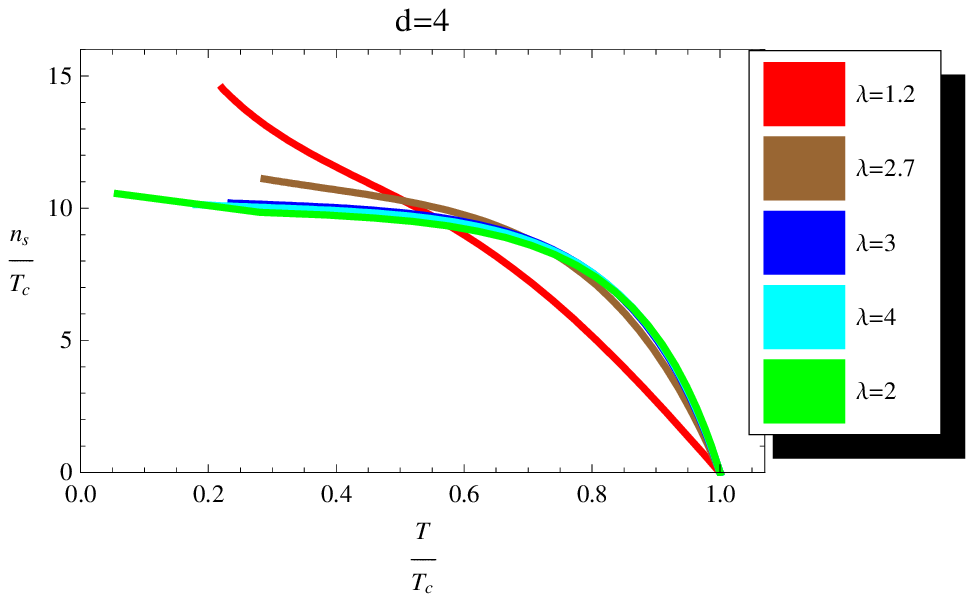}
\caption{Superfluid density below the critical temperature in the boundary theory.}
\label{chap3Figure6}
\end{center}
\end{figure}

In the limit $\omega\rightarrow 0$, the superfluid density $n_{s}$ is defined as  the coefficient of the pole  in the imaginary part of conductivity $\mathcal{I} m[\sigma]=n_{s}/\omega$, where $n_{s}$ is the superfluid density. The results of the superfluid density  computed by solving  equation (\ref{eqb21}) in this limit, is shown in figure \ref{chap3Figure6}.
The vanishing of $n_{s}$ at the critical temperature is in agreement with the Ginzburg-Landau theory.

\begin{figure}[h]
\includegraphics[scale=0.65]{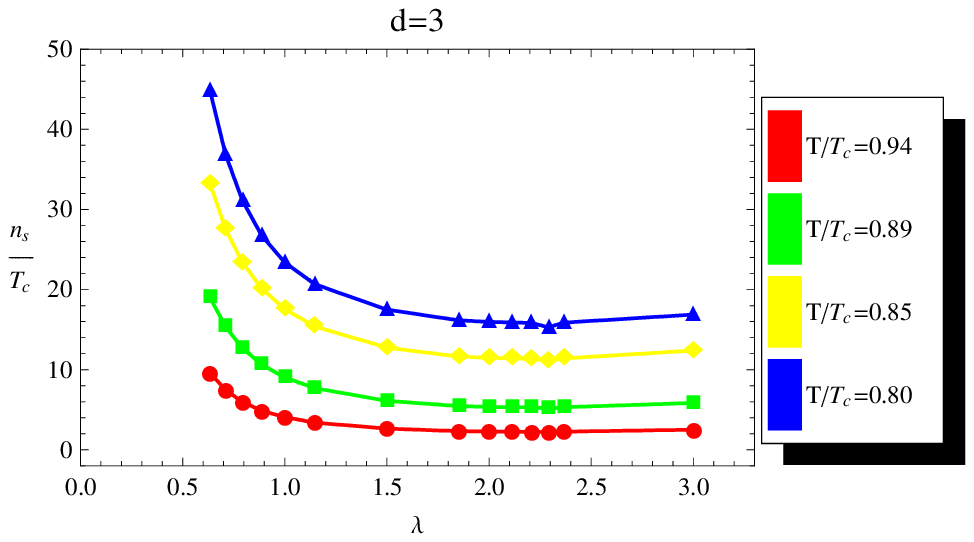}
\includegraphics[scale=0.65]{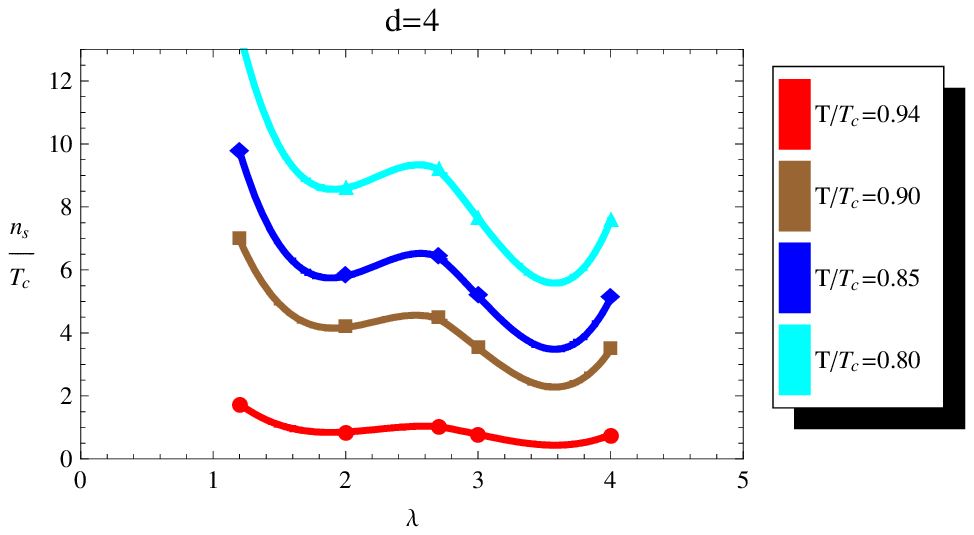}
\caption{Superfluid density as function of the dimension of the dual operator, at  different fixed temperatures}
\label{chap3Figure7}
\end{figure}

The dependence of the  $n_{s}$  on the scaling dimension  calculated at various fixed  temperatures below $T_{c}$ is shown in figure \ref{chap3Figure7}.
Observe that for $\lambda \geq \lambda_{BF}$, $n_s$ is not sensitive to changes in $\lambda$, suggesting that this class of holographic superconductors may not be stable against perturbations by external magnetic field, since $n_{s}$ is related to the current which generate the electromagnetic field if the boundary theory was gauged. 

\begin{figure}[htb!]
\includegraphics[scale=0.65]{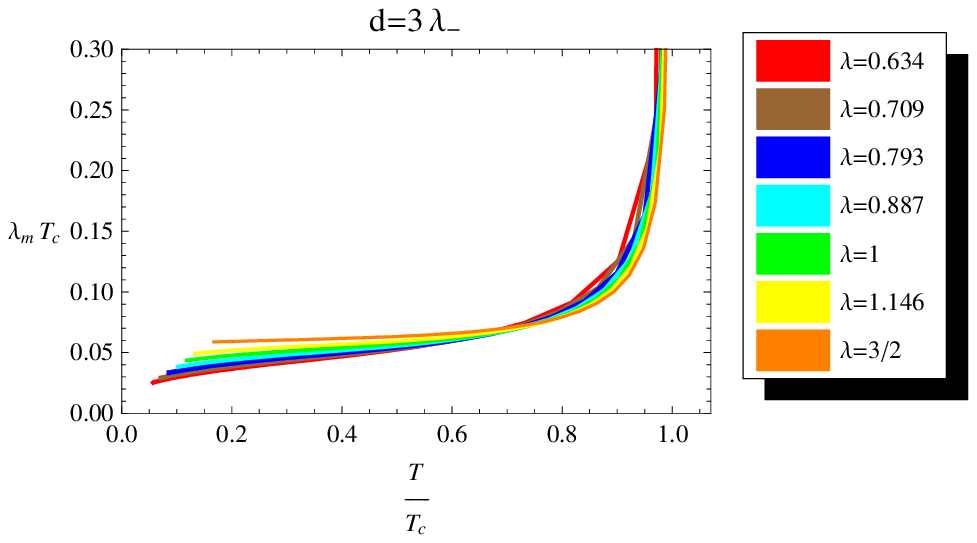}
\includegraphics[scale=0.65]{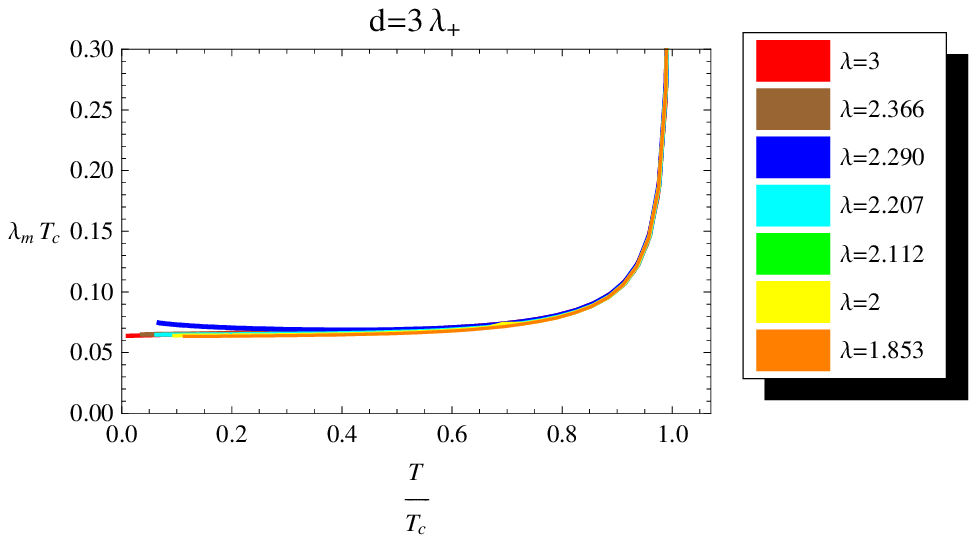}
\begin{center}
\includegraphics[scale=0.65]{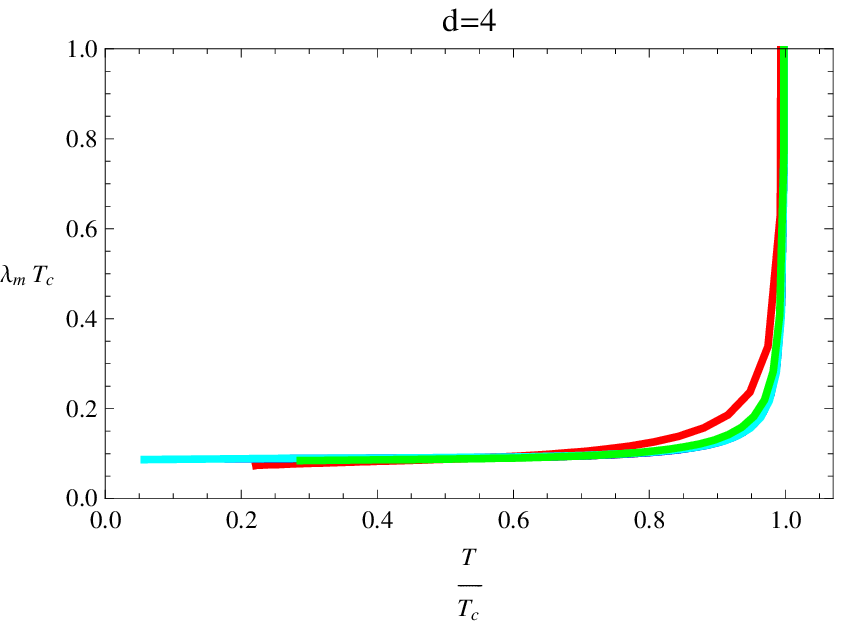}
\caption{The magnetic penetration depth below the critical temperature in the dual field theory.}
\label{pene1}
\end{center}
\end{figure}

\begin{figure}[htb!]
\includegraphics[scale=0.65]{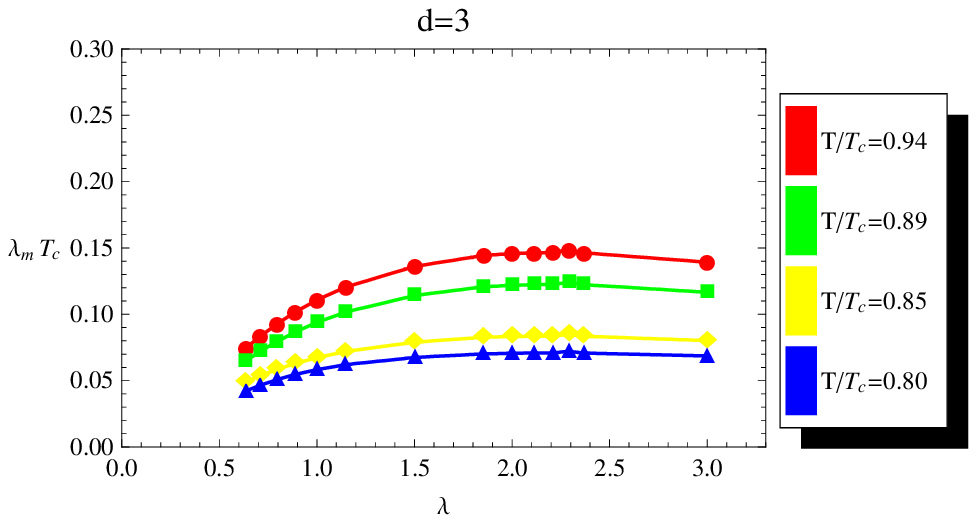}
\includegraphics[scale=0.65]{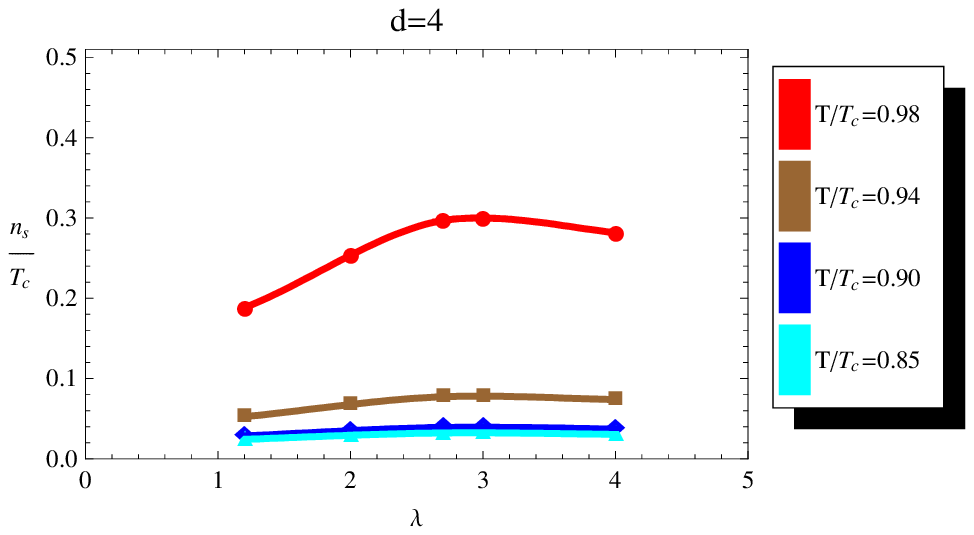}
\caption{Magnetic penetration depth as function of the dimension of the dual operator, at temperatures below $T_{c}$.}
\label{pene2}
\end{figure}

Thus the superfluid density is related to the magnetic penetration depth $\lambda_{m} $ through the first London equation
\begin{equation}\label{London}
 J = -e_{*} n_{s} A,
\end{equation}
where $e_{*}$ is the charge of the order parameter.
Using the Maxwell's equation  for the curl of the magnetic field and assuming that the  current at the boundary can generate its own magnetic field\footnote{i.e weakly gauging the boundary theory as suggested in \cite{Hartnoll:2008kx}.}, the relation between the superfluid density and the magnetic penetration depth appear more explicitly
\begin{eqnarray}\label{magi}
 -\nabla^{2} B = \nabla \times (\nabla \times B) = 4 \pi \nabla \times J = -4\pi n_{s} \nabla \times A = 4\pi n_{s} B \\ \nonumber
\nabla^{2} B = \frac{1}{\lambda_{m}^{2}} B,
\end{eqnarray}
where $\lambda_{m}^{2}=\frac{1}{4\pi n_{s}}$. The  magnetic penetration depth obtained using this relation for both classes of holographic superconductors is shown in  figure \ref{pene1}.
 Notice that the  magnetic penetration depth diverges at $T_{c}$ which is an expected behavior.
Its dependence on the dimension of the dual operator is presented in figure \ref{pene2}

\section{Perturbative Solution}\label{chap3section5}
 At the quantum critical point, equations (\ref{eqb8}) and (\ref{eqb9}) can be solved exactly:
\begin{eqnarray}
 \Psi_{c} =0 ,\\ \nonumber
 \tilde \Phi = q_{c}\left( 1 -\frac{z^{d-2}}{d-2}\right) .
\end{eqnarray}
Other solutions to equations (\ref{eqb8}) and (\ref{eqb9}) can be found in the vicinity of quantum critical point, by a perturbative expansion, since  the superconducting condenstate behaves as (see section \ref{chap3section3}) 
\begin{equation}\label{eqv1}
 \mathcal{\left\langle O\right\rangle } \approx  T_{c}\left( 1-T/T_{c}\right) ^{1/2}
\end{equation}
and vanishes at $T_{c}$.
The results of the numerical calculations in section \ref{chap3section3}  show that at the  critical temperature  $\mu =\rho = q_{c}$\footnote{We use the conventions of \cite{Maeda:2008ir}.}.  
 
Other solutions to equations (\ref{eqb8}) and (\ref{eqb9})  may be obtained to higher order in $\epsilon=\left( 1-T/T_{c}\right)$  and in the manner which still yield the expected fall offs at the AdS boundary.
\begin{eqnarray}\label{eqc6}
\Psi(z)= \epsilon ^{1/2}\Psi_{1}(z) + \epsilon ^{3/2}\Psi_{2}(z) +\epsilon ^{5/2}\Psi_{3}(z)+...\\ \nonumber
\tilde \Phi(z)=\tilde \Phi_{c}(z)+ \epsilon \tilde \Phi_{1}(z)+\epsilon^{2} \tilde \Phi_{2}(z)+...
\end{eqnarray}
Using equation (\ref{eqc6}) in equations (\ref{eqb8}) and (\ref{eqb9}) gives
\begin{equation}\label{eqv2}
\left[ z^{d-1}\frac{d}{dz}\frac{h(z)}{z^{d-1}}\frac{d}{dz}-\frac{m^{2}}{z^{2}}+\frac{\tilde \Phi_{c}^{2}}{h(z)}\right]\Psi_{1} =0,
\end{equation}
\begin{equation}\label{eqv3}
\left[ z^{d-3}h(z)\frac{d}{dz}\frac{1}{z^{d-3}}\frac{d}{dz}\right] \Phi_{1}-\frac{2 \tilde \Phi_{c} \Psi_{1}^{2}}{z^{2}}=0.
\end{equation}
The equations are written in a form most convenient for use in the following analysis. Equation (\ref{eqv2}) decouples from  equation (\ref{eqv3}) to first order in the  perturbative expansion.  We make the  following definitions  for clearer  presentation:
\begin{eqnarray}
\mathcal{L_{\psi}}:=\left[ z^{d-1}\frac{d}{dz}\frac{h(z)}{z^{d-1}}\frac{d}{dz}-\frac{m^{2}}{z^{2}}+\frac{\tilde \Phi_{c}^{2}}{h(z)}\right] \\ \nonumber
\mathcal{L_{\phi}}:=\left[ z^{d-3}h(z)\frac{d}{dz}\frac{1}{z^{d-3}}\frac{d}{dz}\right].
\end{eqnarray}

\subsection{Superconducting coherence length}\label{chap4section4}
The correlation length of the order parameter is related to the  superconducting coherence length $\xi$, which appears as a complex  pole of the static correlation function of the order parameter fluctuation in Fourier space~\cite{Maeda:2008ir}:
\begin{equation}
 \left\langle \mathcal{O}(\vec k)\mathcal{O}(-\vec k)\right\rangle \sim \frac{1}{|\vec k|^2 + 1/\xi^2}.
\end{equation}
Following the technique of AdS/CFT correspondence, this correlation length may be calculated within  the probe approximation by perturbing the Maxwell and scalar fields on a fixed black hole background.
We consider only the linear perturbation, with  fluctuations  of the fields  in the  $x-$direction, in the form
\begin{eqnarray}\label{eqb13}
\delta A_{\mu}\left( z,x\right) dx^{\mu}=\left[ A_{x}\left( z,k\right) dx + A_{y}(z,k)dy +\phi \left( z,k\right) dt\right] e^{ikx},\\ \nonumber
\delta \psi \left( z,x\right)  =\frac{1}{\alpha(T)}\left[ \psi(z,k) + i \tilde{ \psi}(z,k) \right] e^{ikx}.
\end{eqnarray}

Using (\ref{eqb13}) on the perturbed Maxwell and scalar fields give the following eigenvalue equations
\begin{equation}\label{eqb14}
\psi'' + \left( \frac{h'}{h} + \frac{d-1}{z}\right) \psi' - \frac{\tilde k^{2}\psi}{h}+ \frac{\tilde \Phi^{2}\psi}{h^{2}}+\frac{2\tilde \Phi \Psi}{h^{2}}\phi - \frac{m^{2}}{z^{2}h}\psi=0,
\end{equation}
\begin{equation}\label{eqb14b}
\phi'' -\frac{d-3}{z}\phi'- \tilde k^{2}\phi - \frac{2\Psi}{z^{2}}\phi - \frac{4\tilde \Phi \Psi}{h z^{2}}\psi=0,
\end{equation}
\begin{equation}\label{eqb14c}
A''_{y}+ \left[ \frac{h'}{h}-\frac{d-3}{z}\right] A'_{y}- \frac{\tilde k^{2}}{h}A_{y} - \frac{2\Psi^{2}}{z^{2}h}A_{y}= 0,
\end{equation}
where $\tilde k = k/\alpha(T)$. 
Regularity  at the horizon implies that
\begin{eqnarray}\label{eqb15}
\phi=0 \\ \nonumber
\psi '=-\frac{k^{2}\psi}{d z_{+}^{d-1}}-\frac{m^{2}\psi}{d z_{+}^{d+1}}\\ \nonumber
A_{y}'=-\frac{k^{2}A_{y}}{d z_{+}^{d-1}}-\frac{2|\Psi|^{2}A_{y}}{d z_{+}^{d+1}}.
\end{eqnarray}
Analytical treatment is possible  for the eigenvalue equations in the limit  $T\rightarrow T_{c}$~\cite{Maeda:2008ir}.

\begin{figure}[htb!]
\includegraphics[scale=0.65]{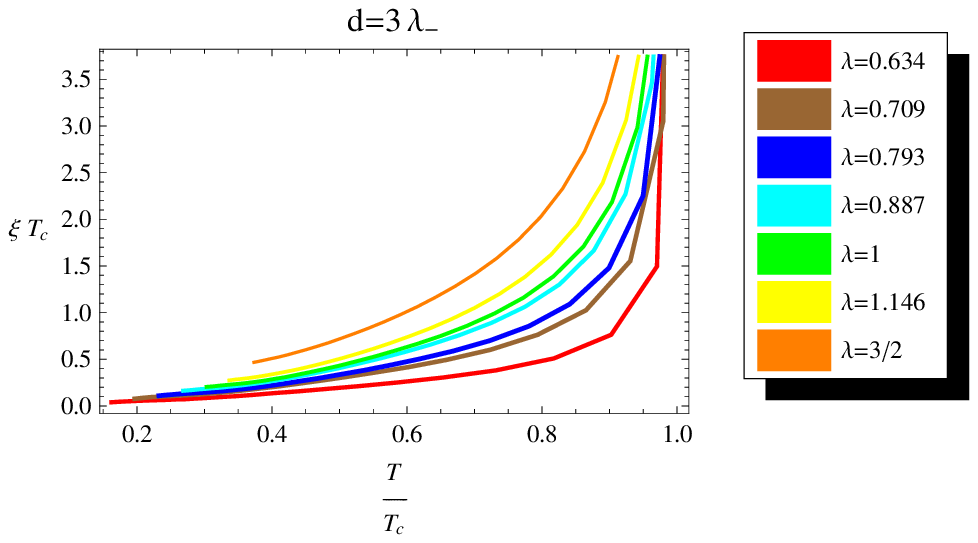}
\includegraphics[scale=0.65]{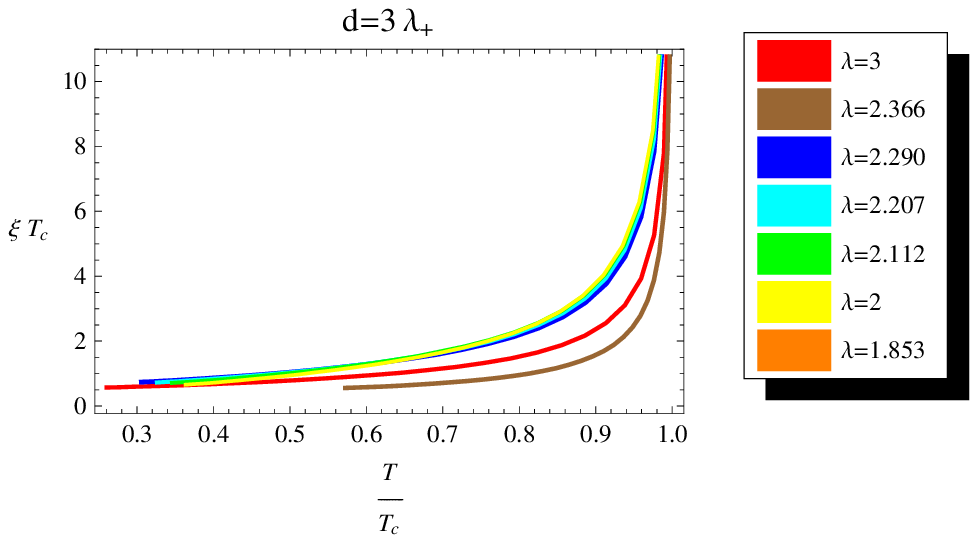}
\begin{center}
 \includegraphics[scale=0.65]{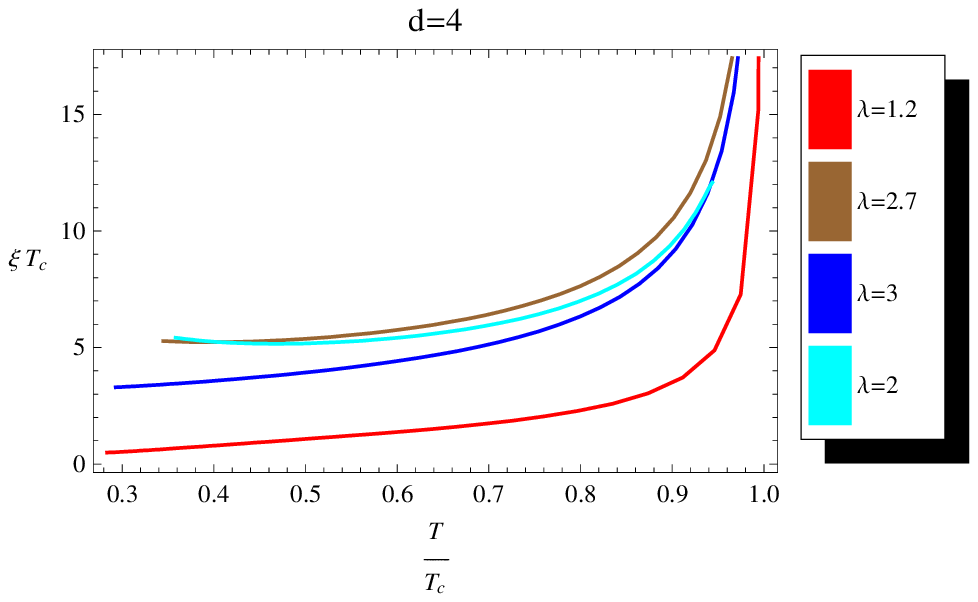}
\end{center}
\caption{Superconducting coherence length of holographic superconductors plotted as a function of temperature.}
\label{chap4Figure8}
\end{figure}

Using the series expansion (\ref{eqc6}) in equations (\ref{eqb14}), (\ref{eqb14b}) and (\ref{eqb14c}) yield

\begin{eqnarray}\label{eqc17}
\mathcal{L_{\psi}}\psi=\tilde{k^{2}}\psi-\frac{2\epsilon \tilde \Phi_{c}\tilde \Phi_{1}}{h(z)}\psi-\frac{2\epsilon^{1/2}\tilde \Phi_{c}\Psi_{1}}{h(z)} \phi \\ \nonumber
\mathcal{L_{\phi}}\phi=\tilde{k^{2}}\phi+\frac{2\epsilon \Psi_{1}^{2}}{z^{2}}\phi+\frac{4\epsilon^{1/2}\tilde \Phi_{c}\Psi_{1}}{z^{2}}\psi.
\end{eqnarray}
The solution to equation (\ref{eqc17}) of interest are those that satisfy the regularity condition at the horizon (\ref{eqb15}) and have an expected fall off at the AdS boundary (equation \ref{eqb16}).
One trivial solution is the zeroth  order solution $\phi_{0}$ and $\psi_{0}$: 
\begin{eqnarray}\label{eqc18b}
\psi_{0}=\Psi_{1} \\ \nonumber
\phi_{0}=0
\end{eqnarray}
Non-trivial solutions can be found by a series expansion around the zeroth order solution in powers of $\epsilon$
\begin{eqnarray}\label{eqc19}
\psi = \Psi_{1} +\epsilon\psi_{1} + \epsilon^{2}\psi_{2}+...\\ \nonumber
\phi = \epsilon^{1/2}\phi_{1} +\epsilon^{3/2}\phi_{2}+...\\ \nonumber
\tilde{k^{2}}=\epsilon \tilde{k^{2}}_{1} +\epsilon^{2} \tilde{k^{2}}_{2}+...
\end{eqnarray}
Substituting the  expansion (\ref{eqc19}) into equation (\ref{eqc17}) yields
\begin{equation}\label{eqc20} 
\mathcal{L_{\psi}}\psi=k_{1}^{2}\Psi_{1}-\frac{2\tilde \Phi_{c}\Psi_{1}}{h(z)}\left( \tilde \Phi_{1}+\phi_{1}\right) 
\end{equation}
\begin{equation}\label{eqc21}
\mathcal{L_{\phi}}\phi_{1}=\frac{4 \tilde \Phi_{c}\Psi^{2}_{1}}{z^{2}}.
\end{equation}
In this approximation it is easy to see that the equations of motion for $\Phi_{1}$ and $\phi_{1}$ only differ by a factor of two.
Equation (\ref{eqc20}) can be solved for $k$ by defining  an inner product for the states $\psi_{1}$ and $\psi_{2}$ which satisfy the boundary condition  at the AdS boundary and is  well behaved  at the horizon (\ref{eqb15}).
\begin{equation}\label{eqc22}
\left\langle \psi_{I}\psi_{II}\right\rangle =\int_{0}^{1}\frac{dz}{z^{d-1}}\psi^{*}_{I}\psi_{II}.
\end{equation}
Because  $\mathcal{L_{\psi}}$  is hermitian for non-zero negative mass squared, taking the inner product of equation (\ref{eqc20}) gives 
\begin{equation}\label{eqc23}
\left\langle \Psi_{1}|\mathcal{L_{\psi}}|\psi_{1}\right\rangle =k_{1}^{2}\left\langle \Psi_{1}|\Psi_{1}\right\rangle -\left\langle \Psi_{1}\frac{2\tilde \Phi_{c}\Psi_{1}}{h(z)}\left( \tilde \Phi_{1}+ \phi_{1}\right)\right\rangle 
\end{equation}
Using the inner product (\ref{eqc22}) and the constraint $ \mathcal{L_{\psi}}\Psi_{1}=0$ in  equation (\ref{eqc23}) we obtain
\begin{equation}\label{eqc23b}
k_{1}^{2}\left\langle \Psi_{1}|\Psi_{1}\right\rangle=\left\langle \Psi_{1}\frac{2\tilde \Phi_{c}\Psi_{1}}{h(z)} \tilde \Phi_{1}\right\rangle +2\int_{0}^{1}dz\frac{\tilde \Phi_{c}\Psi_{1}^{2}}{z^{d-1}h(z)}\phi_{1}.
\end{equation}
Equation (\ref{eqc23b}) may be simplified by considering the equation of motion for the mode $\Psi_{2}$:
\begin{equation}\label{eqc24}
\mathcal{L_{\psi}}\Psi_{2}=\frac{2\tilde \Phi_{c}\Psi_{1}}{h(z)}\tilde \Phi_{1}.
\end{equation} 
Since   equation (\ref{eqc22}) is well defined for $\Psi_{1}$  and  it is hermitian, the first term in  the right hand side  of equation (\ref{eqc23b}) is  zero.
 Using equation (\ref{eqc22}) and  $\tilde k^{2}=\epsilon \tilde k_{1}^{2}$, the eigenvalue $\tilde k$  in a  first order approximation may be written as.
 \begin{equation}\label{eqc25}
 \tilde k^{2}=\epsilon\frac{N}{D} + \mathcal{O}(\epsilon^{2})
 \end{equation}
 where
 \begin{eqnarray}\label{eqc26}
 N=2\int_{0}^{1}dz\frac{\tilde \Phi_{c}\Psi_{1}^{2}}{z^{d-1}h(z)}\phi_{1}\\ \nonumber
 D=\int_{0}^{1}\frac{\Psi_{1}^{2}}{z^{d-1}}
 \end{eqnarray}
 This result was first derived in~\cite{Maeda:2008ir} for $m^2=-2$, and it  is  shown to  hold for all the masses that satisfy the unitarity bound in $d-$dimensions, except for  $d=2$ where the  scalar potential diverges.
Now the superconducting coherence length is given by
\begin{equation}\label{eqc27}
\xi = \frac{\epsilon^{-1/2}}{\alpha(T)}\sqrt{\frac{D}{N}} +\mathcal{O}(\epsilon^{2})
\end{equation}
Figure \ref{chap4Figure8} shows the results obtained from calculating  the $\xi$ using equation (\ref{eqc27}) for various condensates. We have used the  boundary conditions obtained in section \ref{chap3section3} to solve for  $\Psi_{1}$  and $\phi_{1} $.

\begin{figure}[htb!]
\includegraphics[scale=0.65]{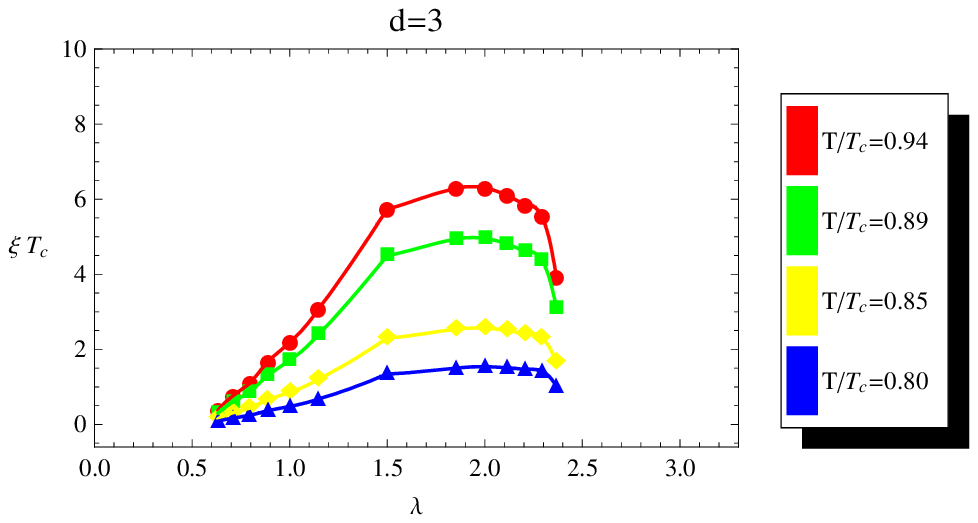}
\includegraphics[scale=0.65]{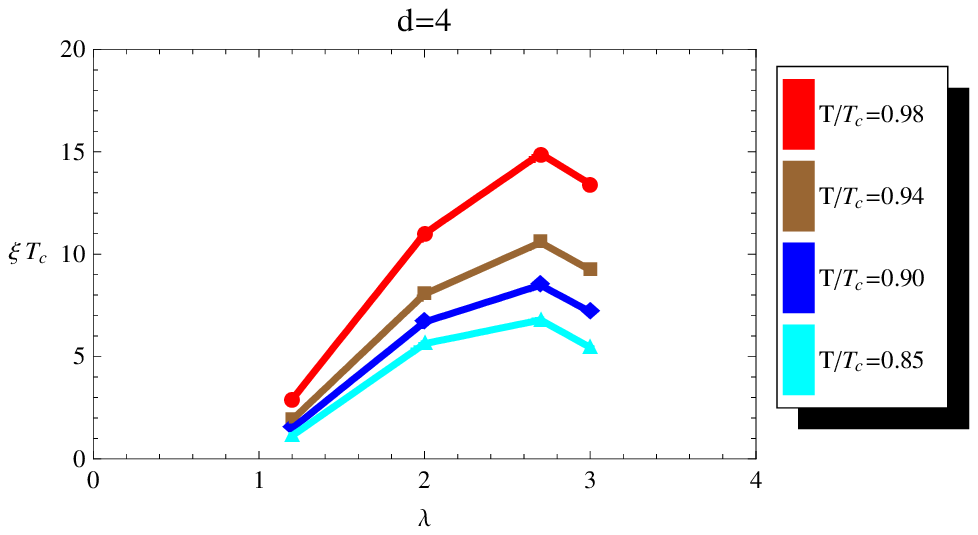}
\caption{Superconducting correlation length  as a function $\lambda$ }
\label{chap4Figure9}
\end{figure}

The numerical accuracy becomes very unsatisfactory for $m^2=0$. As a result, we did not include it in the figure \ref{chap4Figure8}.
The dependence of the superconducting correlation length on the scaling dimensions of the dual condensates is shown in figure \ref{chap4Figure9}.

\subsection{Magnetic penetration depth}\label{chap4section5}
As stated in section \ref{chap3section4}, the magnetic penetration  depth may be calculated from the London current.  This can also be calculated by solving equation \ref{eqb14c} perturbatively in the limit $T\rightarrow T_{c}$ at zero frequency and momentum. The relevant portion of the Maxwell's equation is given by
\begin{equation}\label{eqd2}
z^{d-3}\frac{d}{dz}\frac{h}{z^{d-3}}\frac{dA_{y}}{dz}-\frac{2\Psi^{2}}{z^{2}}A_{y}=0.
\end{equation}
The Maxwell field can be expanded as  $A=A_{0}+\epsilon A_{1}$ in the neighborhood of the QCP, which leads to the following equations
\begin{equation}\label{eqd3}
\frac{d}{dz}\frac{h}{z^{d-3}}\frac{d}{dz}A_{0}=0
\end{equation}
\begin{equation}\label{eqd4}
\frac{d}{dz}\frac{h}{z^{d-3}}\frac{d}{dz}A_{1}-\frac{2\Psi_{1}}{z^{d-1}}A_{0}=0,
\end{equation}
where the subscript,$y$, has been dropped for clarity. One of the solutions to equation (\ref{eqd3}), which  satisfies the required boundary conditions is
\begin{equation}
A_{0}= C ,
\end{equation}
where $C$ is a constant.
 Hence the first order mode becomes
 \begin{equation}\label{eqd5}
 \frac{dA_{1}}{dz}=- \frac{2A_{0}z^{d-3}}{h(z)}\int_{z_{0}^{1}}dz_{0}\frac{|\Psi_{1}(z_{0})|^{2}}{z^{d-1}_{0}}
 \end{equation}
 Integrating this expression (\ref{eqd5}) yields
 \begin{equation}\label{eqd6}
 A_{1}(z)= A_{0}-2A_{0}\int_{0}^{1}dz\frac{z^{d-3}}{h(z)}\int_{z_{0}}^{1}dz_{0}\frac{|\Psi_{1}(z_{0})|^{2}}{z^{d-1}_{0}} + \mathcal{O}(\epsilon^{2})
 \end{equation}
 Here  $A_{0}$ is the  constant of integration. Using $A=A_{0}+\epsilon A_{1}$
 \begin{equation}\label{eqd7}
 A(z)=A_{0} - 2\epsilon A_{0}\int_{z}^{1}dz\frac{z^{d-3}}{h(z)}\int_{z_{0}}^{1}dz_{0}\frac{|\Psi_{1}(z)|^{2}}{z^{d-1}_{0}} + \mathcal{O}(\epsilon^{2}).
 \end{equation}
 Near the boundary $h(z)\approx 1$
 \begin{equation}\label{eqd8}
 A(z) = A_{0} - 2\epsilon A_{0}\int_{z}^{1}dz z^{d-3}\int_{z_{0}}^{1}dz\frac{|\Psi_{1}(z_{0})|^{2}}{z^{d-1}_{0}} + \mathcal{O}(\epsilon^{2}).
 \end{equation}
 From the dictionary of AdS/CFT correspondence, the current is identified as
 \begin{equation}\label{eqd9}
\left\langle j\right\rangle =-\frac{1}{\kappa^{2}_{d}}\left( \frac{4\pi T_{c}}{d (d-2)}\right) \epsilon \int_{0}^{1}dz\frac{\Psi_{1}^{2}}{z^{d-1}}A_{0}(x)+\mathcal{O}(\epsilon^{2}),
\end{equation}
and, for $\epsilon=(1-T/T_{c})$, the current becomes 
\begin{equation}\label{eqd10}
\left\langle j\right\rangle =-\frac{1}{\kappa^{2}_{d}}\left( \frac{4\pi T_{c}}{d(d-2)}\right) (1-T/T_{c}) \int_{0}^{1}dz\frac{\Psi_{1}^{2}}{z^{d-1}}A_{0}+\mathcal{O}(\epsilon^{2})
\end{equation}

\begin{figure}[htb!]
\includegraphics[scale=0.65]{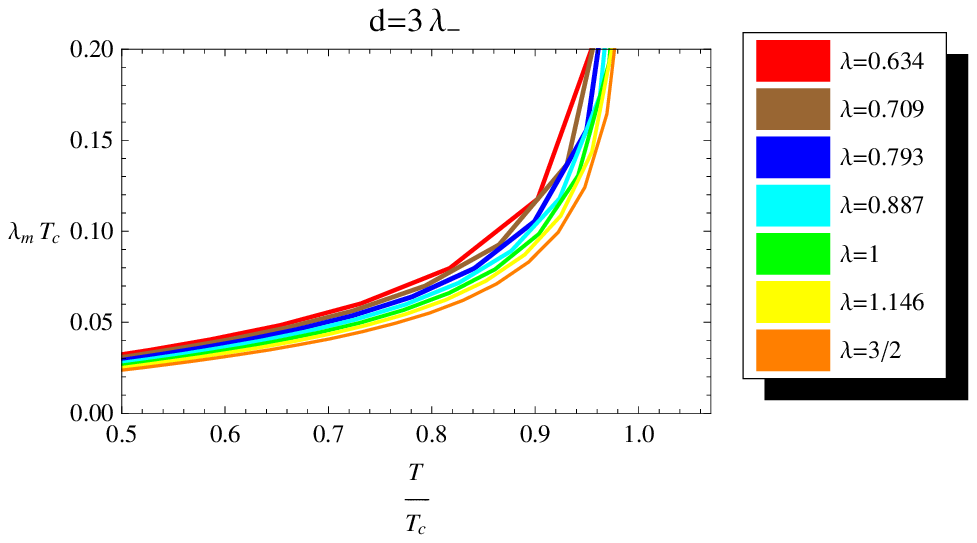}
\includegraphics[scale=0.65]{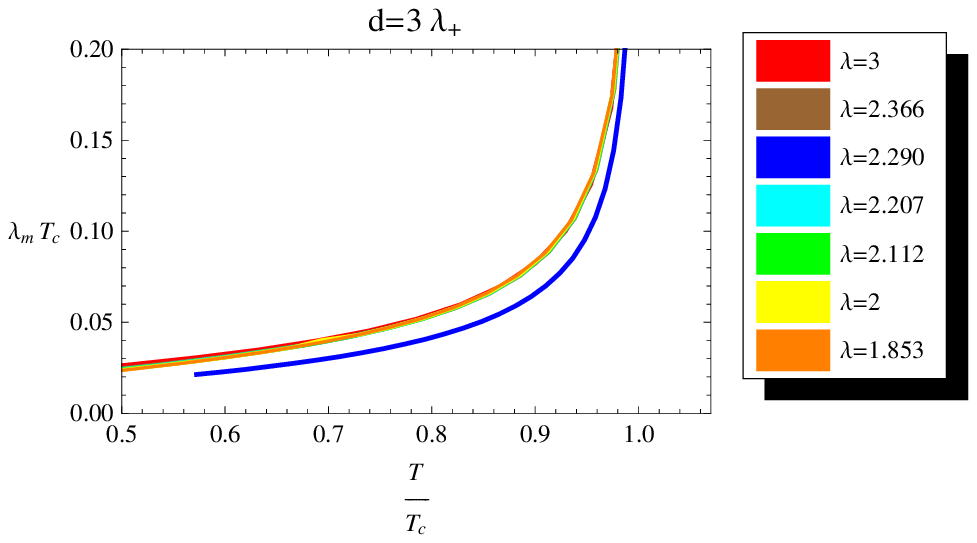}
\begin{center}
 \includegraphics[scale=0.65]{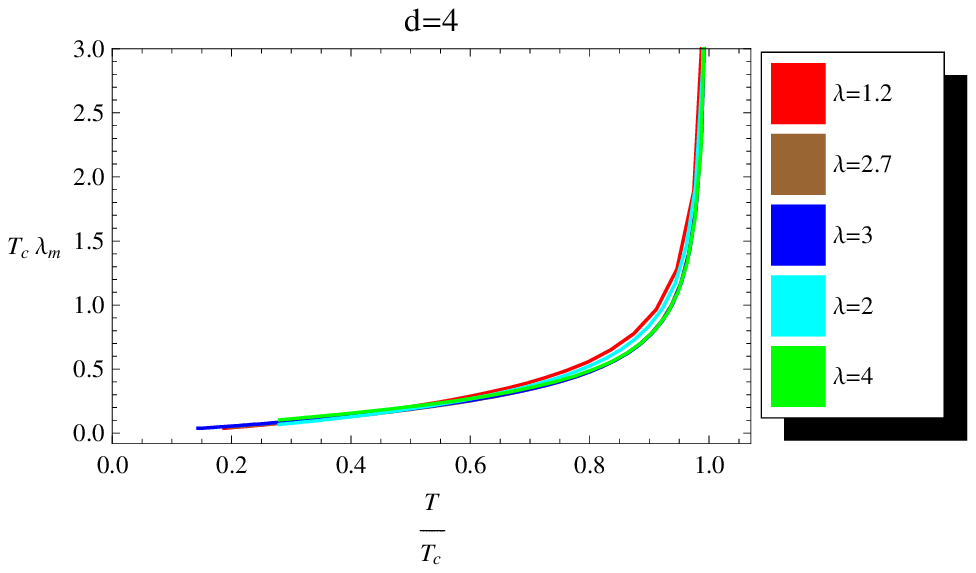}
\end{center}
\caption{Magnetic penetration depth below the critical temperature in the superconducting phase. }
\label{chap4Figure10}
\end{figure}

\begin{figure}[htb!]
\includegraphics[scale=0.65]{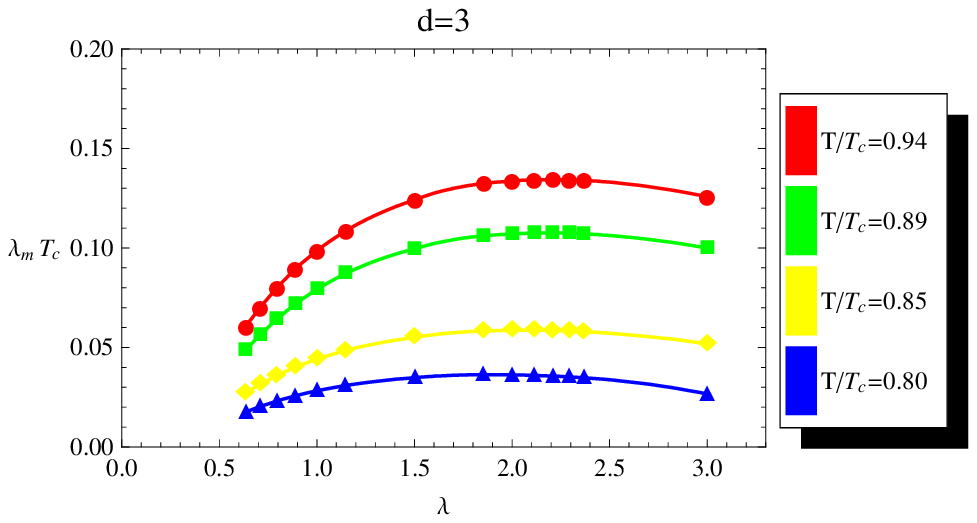}
\includegraphics[scale=0.65]{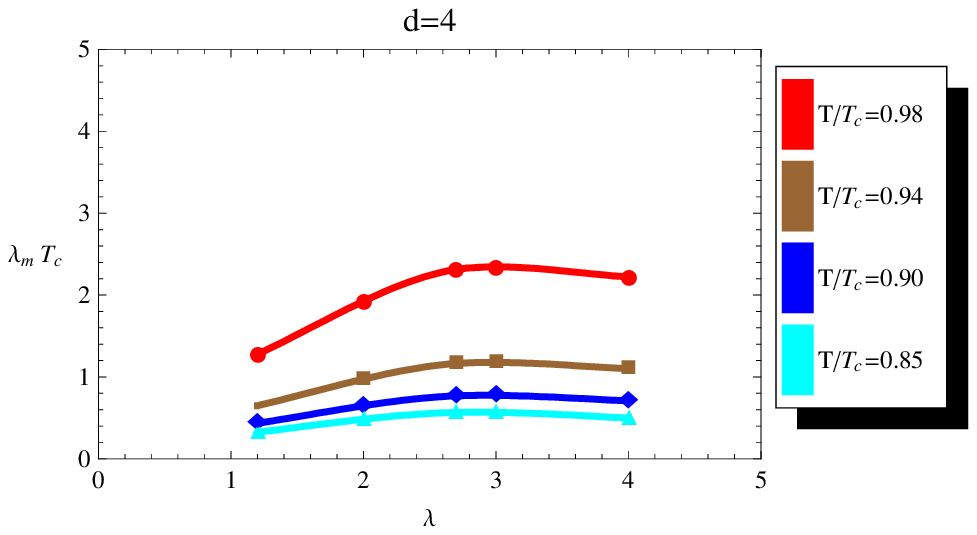}
\caption{  Magnetic penetration depth as a function $\lambda$. }
\label{chap4Figure11}
\end{figure}

The magnetic penetration depth is then defined (see equation \ref{magi}) as
\begin{equation}\label{eqd11}
\lambda_{m}=\sqrt{\left[\frac{1}{\kappa^{2}_{d}}\left( \frac{4\pi T_{c}}{d(d-2)}\right) (1-T/T_{c}) \int_{0}^{1}dz\frac{\Psi_{1}^{2}}{z^{d-1}}\right] ^{-1}  }
\end{equation}
Using equation (\ref{eqc27}) and (\ref{eqd11}), the Ginzburg-Landau parameter becomes
\begin{equation}\label{ginz}
\kappa=\frac{\lambda_{m}}{\xi}.
\end{equation}
To solve for $\lambda_{m}$ we use the relation $\Psi = \epsilon^{1/2}\Psi_{1} +\mathcal{O}(\epsilon)$ to compute $\Psi$ instead of $\Psi_{1}$. This offers some numerical simplification. The results of the numerical computations are presented in figure \ref{chap4Figure10}.
The dependence of the magnetic penetration depth $\lambda_{m}$ on the scaling  dimensions  of the dual condensates  is shown in figure \ref{chap4Figure11}. 

Observe that the results of the magnetic penetration depth, calculated using a perturbative approach and the one calculated from superfluid density are in agreement. This agreement show that the perturbative treatment captures the physics of interest in the vicinity of the QCP.

\begin{figure}[htb!]
\includegraphics[scale=0.65]{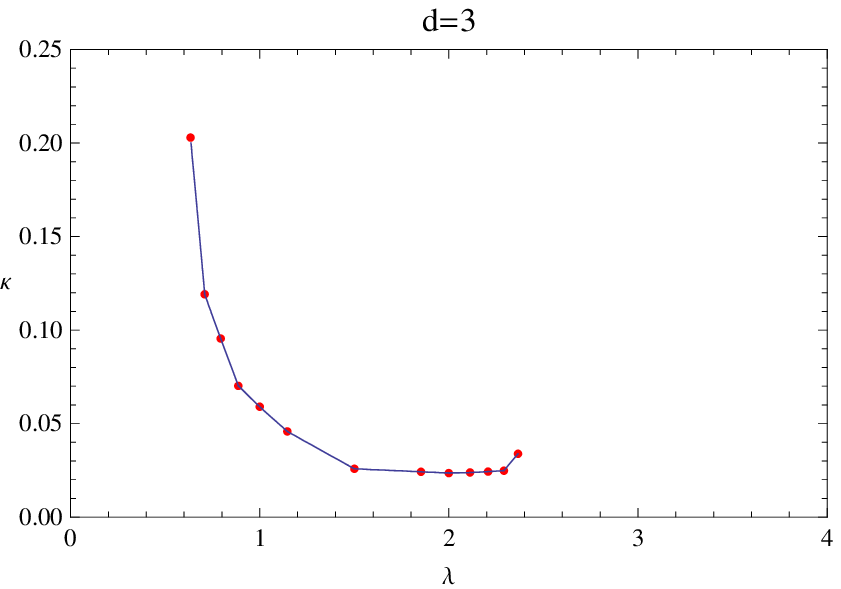}
\includegraphics[scale=0.65]{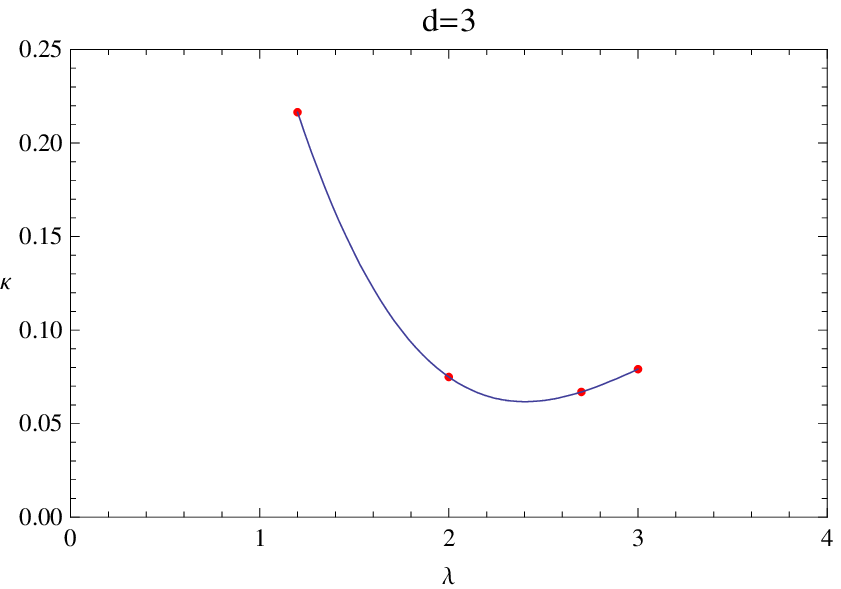}
\caption{Ginzburg-Landau parameter against $\lambda$.}
\label{chap4Figure12}
\end{figure}

The Ginzburg-Landau parameter $\kappa =\lambda_{m}/\xi$ can be calculated from equations  (\ref{eqc27}) and (\ref{eqd11}). The results  obtained are plotted in figure \ref{chap4Figure12} against the dimension of the dual condensate. 

In Ginzburg-Landau theory, the coefficient $\kappa$ classifies superconductors into two types, i.e $\kappa < 1\sqrt{2}$  for type I superconductors and $\kappa > 1\sqrt{2}$  for type II superconductors. If our boundary theory was gauged, the results in figure \ref{chap4Figure12}  show that at $\lambda =\lambda_{BF}$, there is a change in the relative size of $\kappa$. An obvious interpretation is that for $\lambda <\lambda_{BF}$  superconducting condensates are  of type II, while for $\lambda >\lambda_{BF}$ they are of  type I.  It is interesting to see that similar clear distinction  also exist for holographic superconductors. Although, we should note that the London current also depends on $q$, which was scaled away in the probe limit. The effect of large but finite $q$ is to ensure that $\lambda_{m}$ is greater than $\xi$, i.e the condensate must be type II. Despite being large, there are still indications that a holographic superconductor can be type I. This result is in agreement with   Maeda et. al.~\cite{Maeda:2008ir}, who suggested that holographic superconductors which have  low critical temperature are type I. But Hartnoll et. al.~\cite{Hartnoll:2008kx} showed that holographic superconductor corresponding to dimension one operator, which they studied with high accuracy is a type II. These results are not in any way contradicting, as we have seen that both deductions are correct limits of the  larger class of condensates considered here.


\section{Conclusion}\label{conclude}
We have studied the dependence of  various  physical quantities  associated with the  holographic model of superconductivity on the scaling dimensions of the dual condensates in the ($2+1$) and ($3+1$)-dimensional boundary  theories. Each of these physical quantities  was calculated at a fixed temperature, but for different values of mass squared  $m^2$ (varied in 0.5 unit intervals) in $d=3$ and $d=4$ bulk spacetime dimensions. We considered mainly bulk scalar fields which have  normalizable fall-offs at the AdS boundary. 
The results of this indicate that, there are two distinct superconducting condensates dual to the two modes of scalar field, which have different fall-off behaviors at the AdS boundary. The amount of the condensate dual to the bulk scalar field with slower fall-off $\Psi_{\lambda_{-}}$  converges,  before diverging collectively near zero temperature. Its superconducting phase  is  different from that of the scalar fields with a faster fall-off $\Psi_{\lambda_{+}} $. Certain features indicating a discontinuity  in the  amount of condensates were observed between condensates of the class $\Psi_{\lambda_{-}} $ and those of  the class $ \Psi_{\lambda_{+}}$ at $\lambda=\lambda_{BF}$.  This discontinuity  distinguishes between the two classes.  The Ginzburg-Landau parameter $\kappa$, obtained from the  superconducting coherence length $\xi$ and magnetic penetration depth $\lambda_{m}$, indicates that there is a critical scaling dimension $\lambda_{crit}$  at which the holographic superconductors change from type II to type I. Type I holographic superconductors have very low critical temperatures, unlike those of  type II, which have relatively high critical temperatures.

 It would be very interesting to extend the computations presented in this paper to include the effects of the backreaction of the scalar field on the gravitational background. This would enable us to understand the source of the  divergence for the condensates of the class $\Psi_{\lambda_{-}}$. A treatment involving a complete backreacted geometry would shed some light on the class of condensate that would be associated with the vortex and droplets solutions found in~\cite{Albash:2009ix,Albash:2009iq}. Based on an understanding of real superconductors, one would not expect a type I holographic superconductor to support a stable vortex solution. 
 One might also repeat the  analysis presented here for the action, involving a matter field  considered in~\cite{Franco:2009yz}. This would indicate whether the  features observed here are general, and might apply to an entire class of theories with gravity duals. 
\section*{Acknowledgement}
\vspace{1mm}
 It is a great pleasure to thank Sean Hartnoll and Jeff Murugan for many discussions and special assistance. I  benefited from discussions with  Julian Sonner, Palleb Basu, Petja Salmi and Matthew Roberts. I thank  Alex Hamilton and Andrea Prinsloo for technical assistance on the draft. I acknowledge the financial assistance of NASSP, South Africa, in the form of scholarship, towards this research.

\appendix
 
\section{Conductivity in ($2+1$)-dimensional Boundary Theory ($\lambda_{-}$)}\label{appA}
\begin{figure}[htb!]
 \includegraphics[scale=0.3]{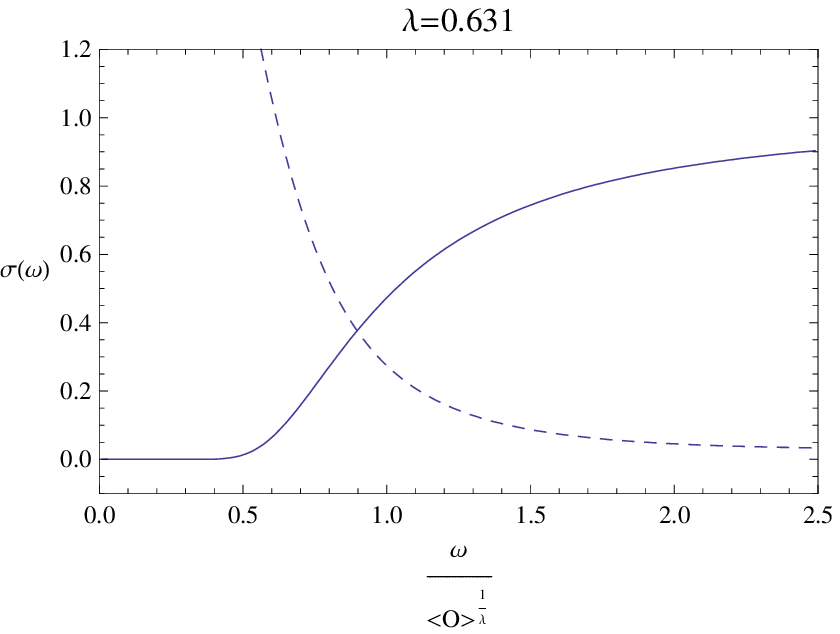}
\includegraphics[scale=0.3]{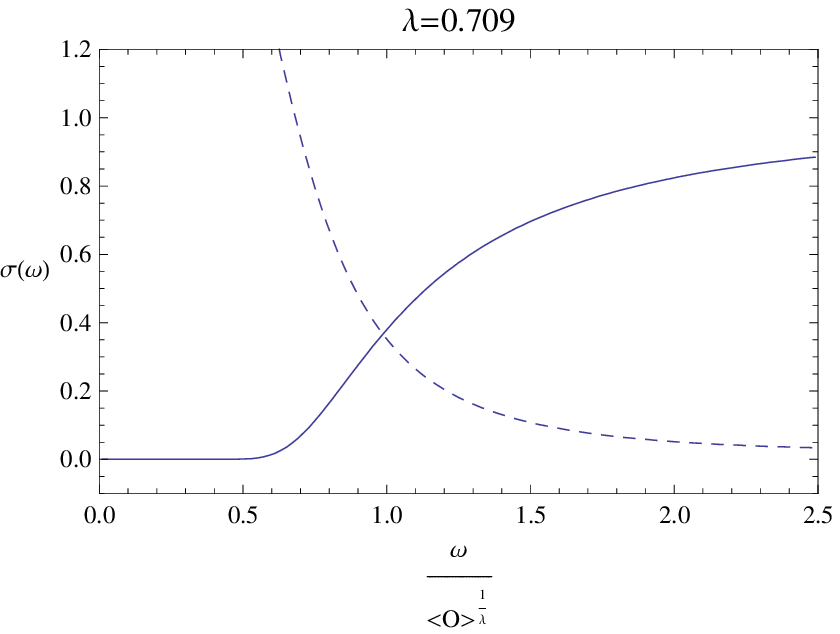}
\includegraphics[scale=0.3]{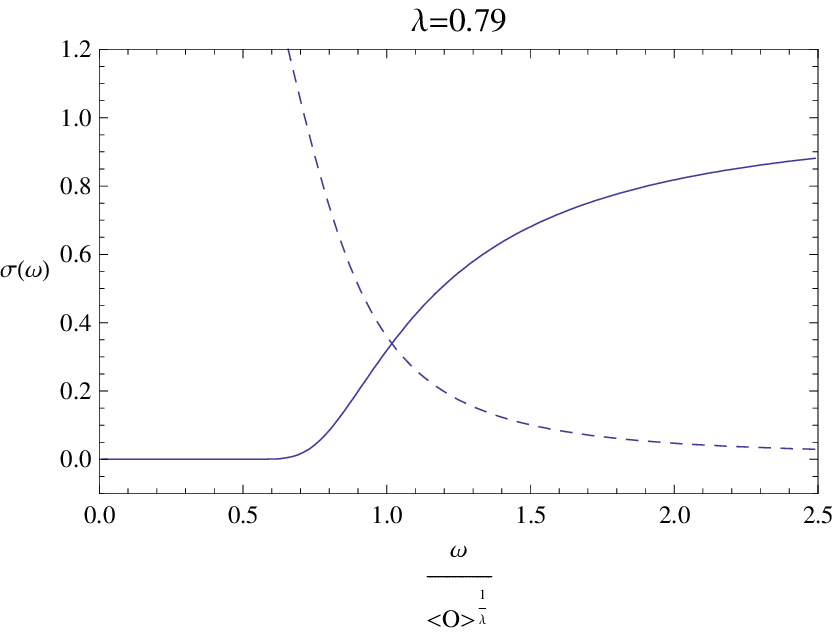}
\includegraphics[scale=0.3]{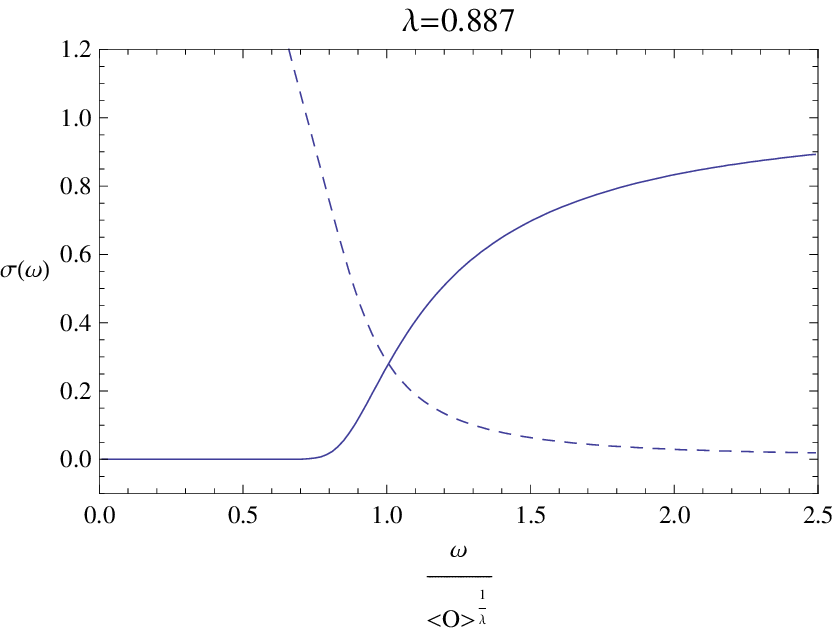}
\includegraphics[scale=0.3]{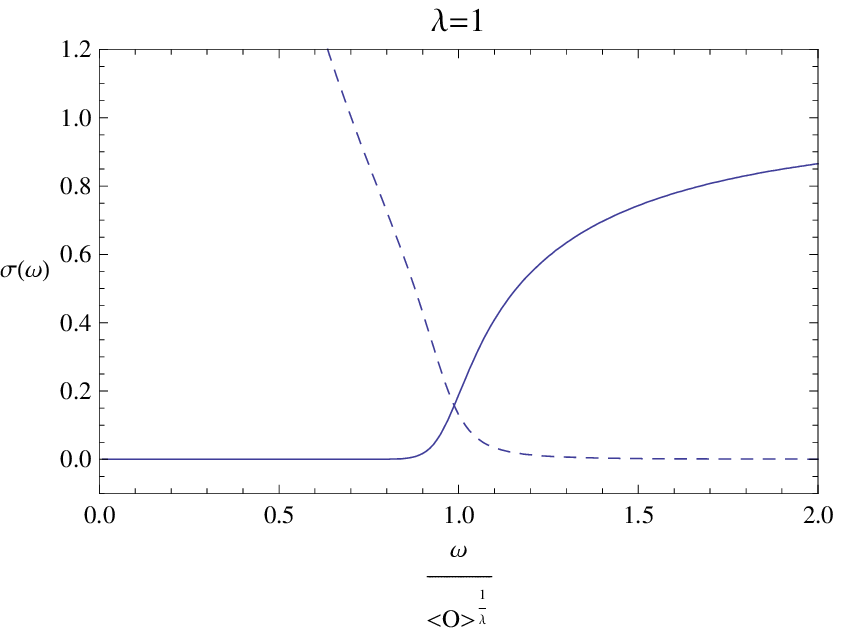}
\includegraphics[scale=0.3]{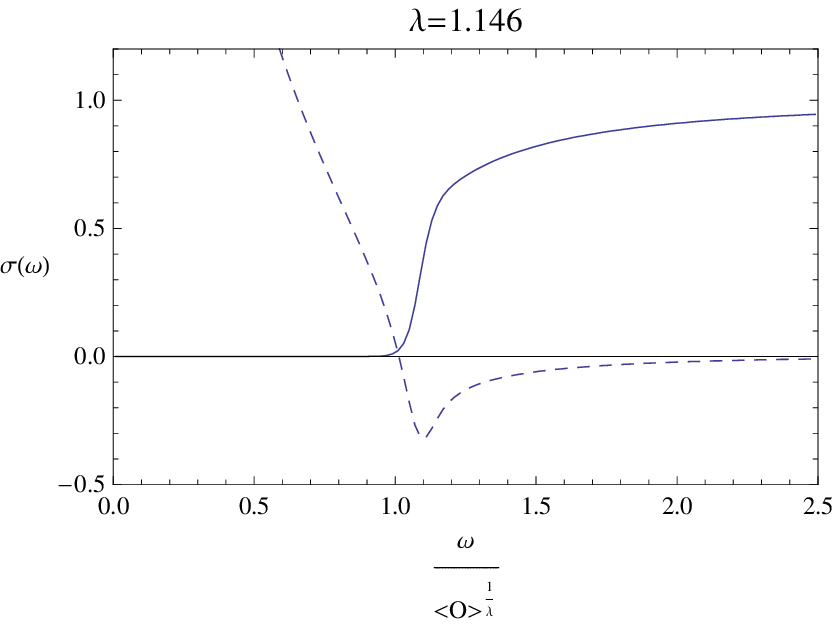}
\includegraphics[scale=0.3]{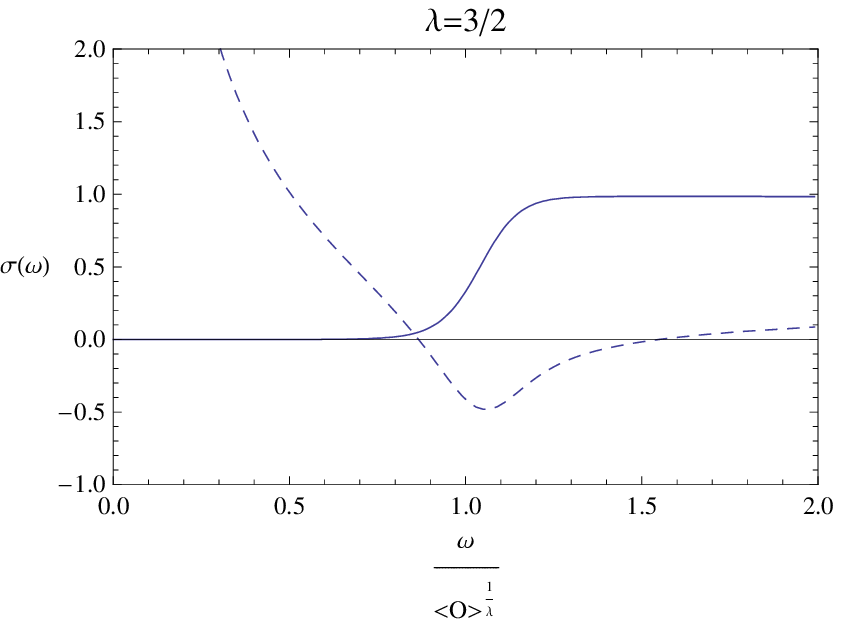}
\caption{Plots of frequency dependent conductivity for condensates of class $\Psi_{\lambda_{-}}$. The frequency is  normalized by the condensate in the superconducting phase. The plots are labelled by the dimension of the operator in the dual field theory.}
\label{Figure4c}
\end{figure}
\section{Conductivity in ($2+1$)-dimensional Boundary Theory ($\lambda_{+}$) }\label{appC}
\begin{figure}[htb]
 \includegraphics[scale=0.3]{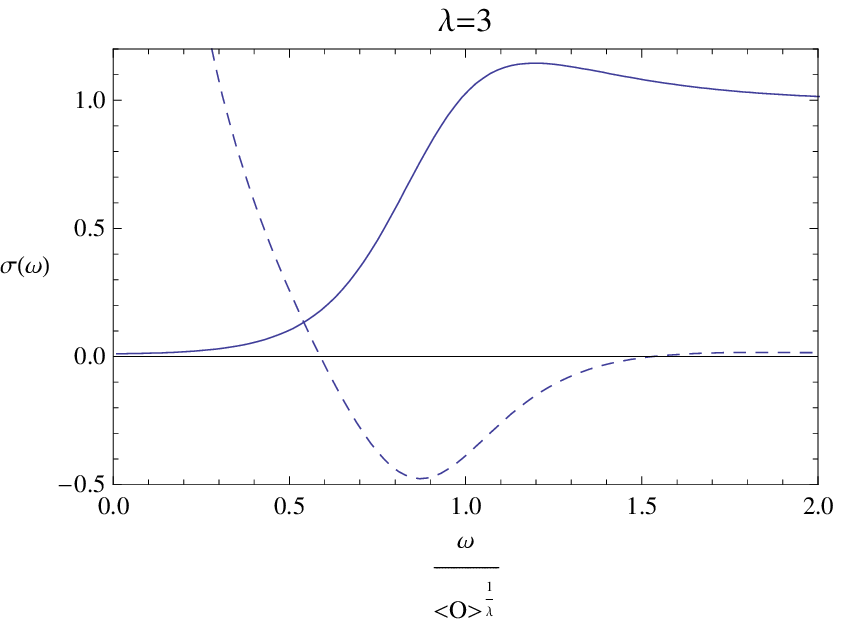}
\includegraphics[scale=0.3]{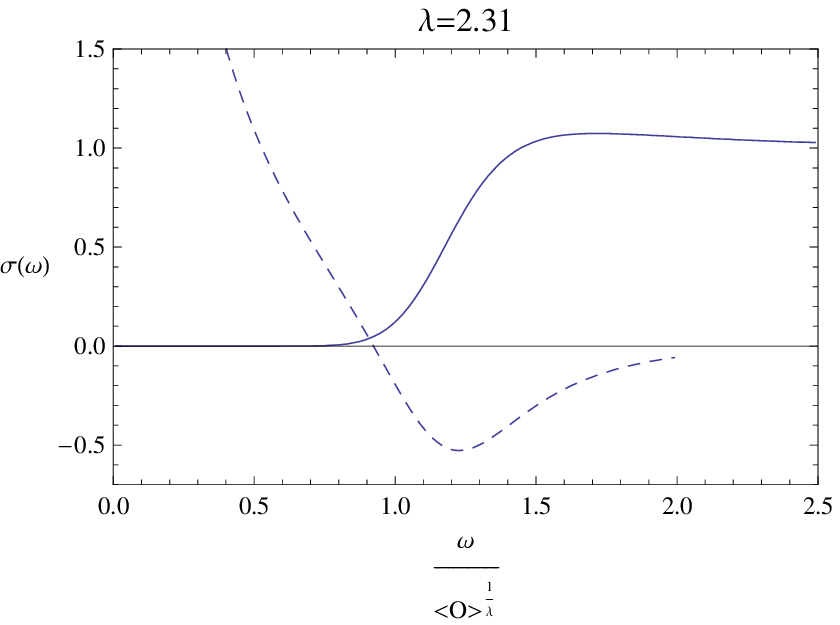}
\includegraphics[scale=0.3]{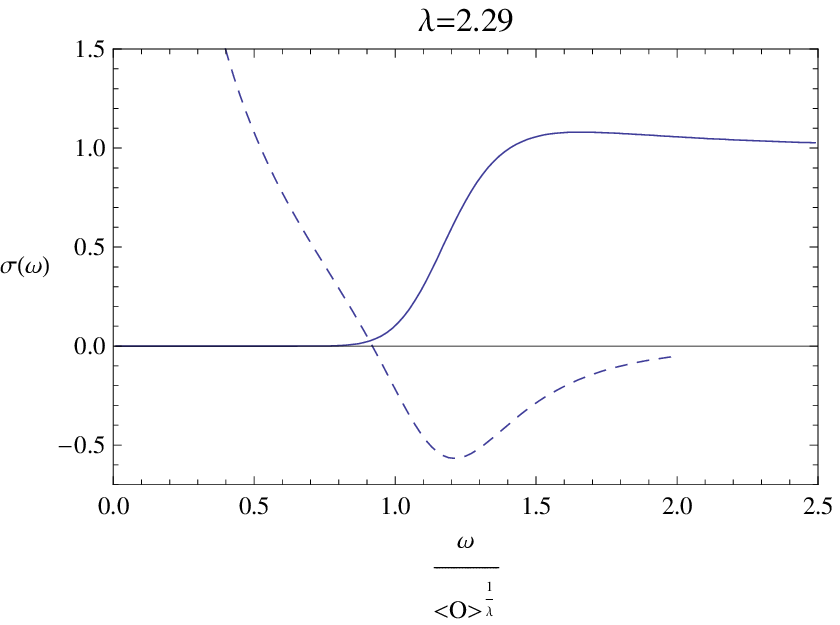}
\includegraphics[scale=0.3]{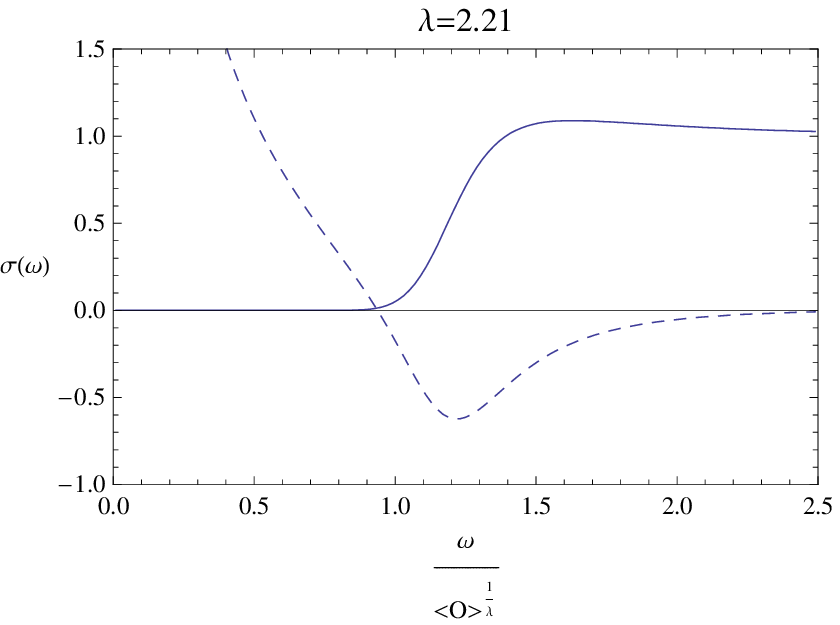}
\includegraphics[scale=0.3]{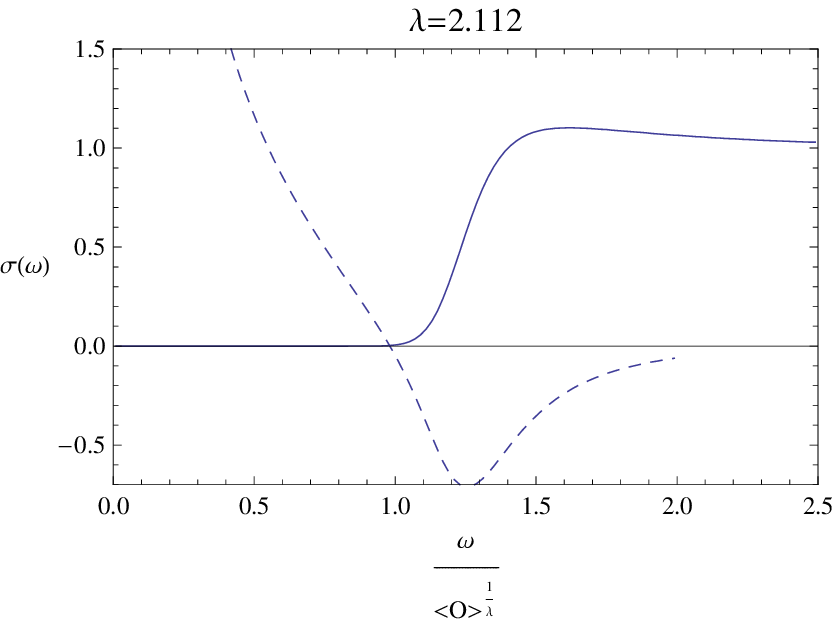}
\includegraphics[scale=0.3]{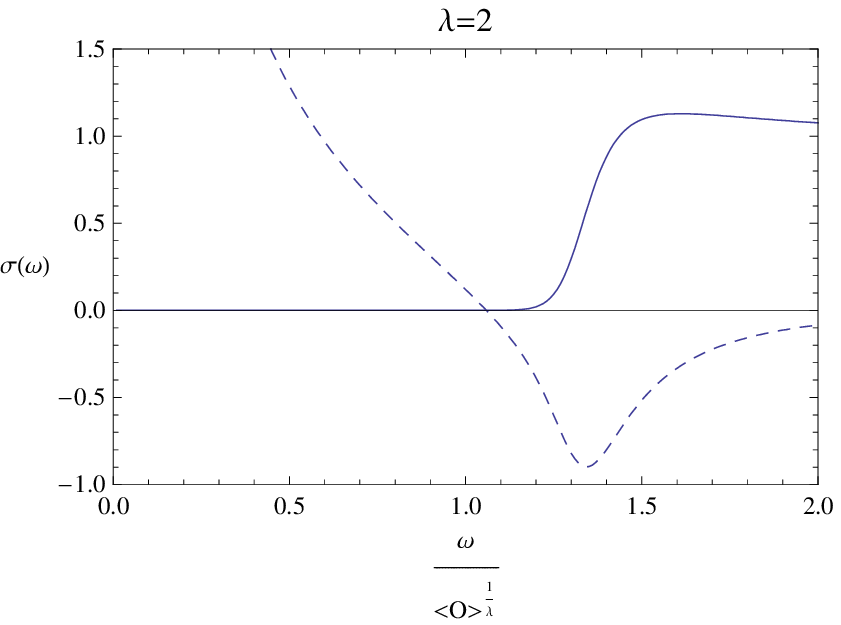}
\includegraphics[scale=0.3]{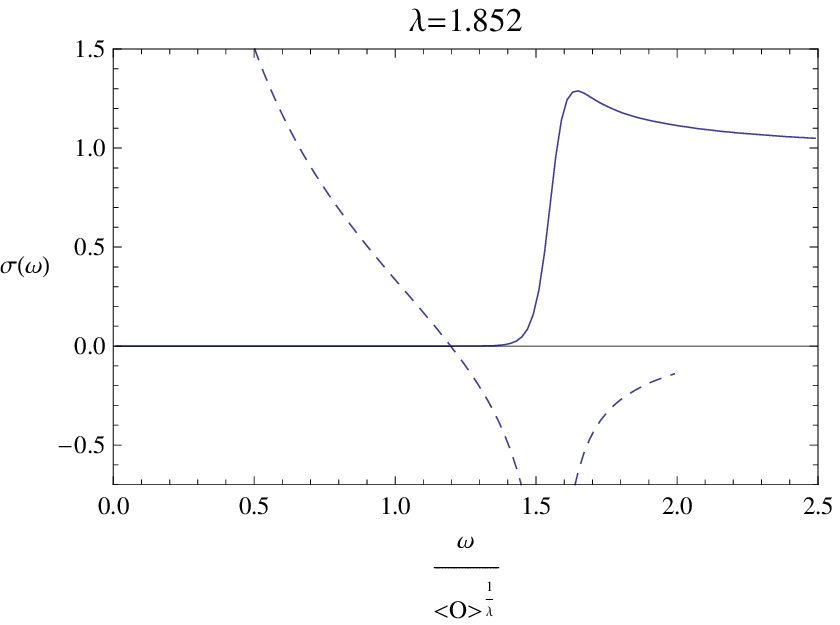}
\caption{Plots of the frequency dependent conductivity for condensates of class $\Psi_{\lambda_{+}}$. The frequency is  normalized by the condensate. The plots are labelled by the dimension of the operator in the boundary theory}
\label{Figure4d}
\end{figure}
\section{Conductivity in ($3+1$)-dimensional Dual Field Theory}\label{appB}
\begin{figure}[htb!]
 \includegraphics[scale=0.3]{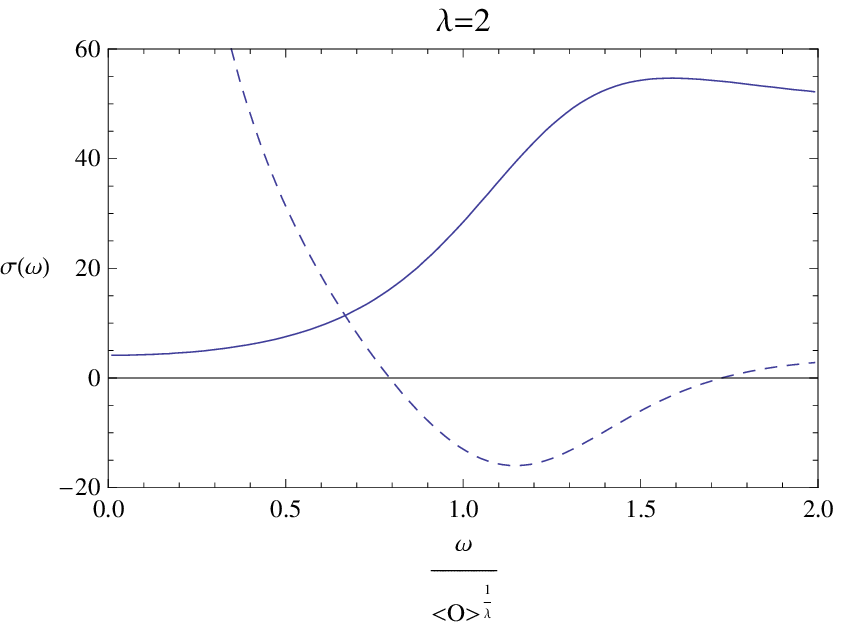}
\includegraphics[scale=0.3]{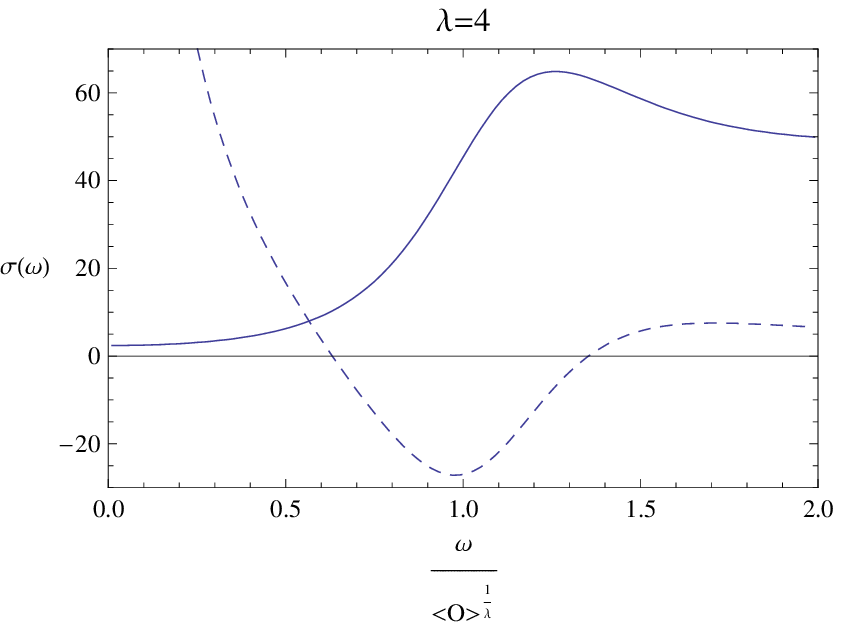}
\includegraphics[scale=0.3]{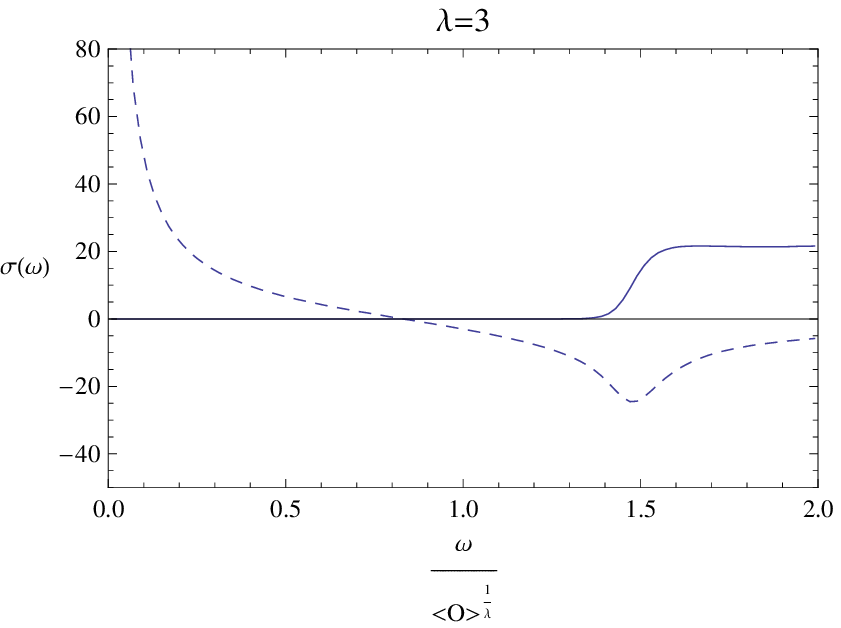}
\includegraphics[scale=0.3]{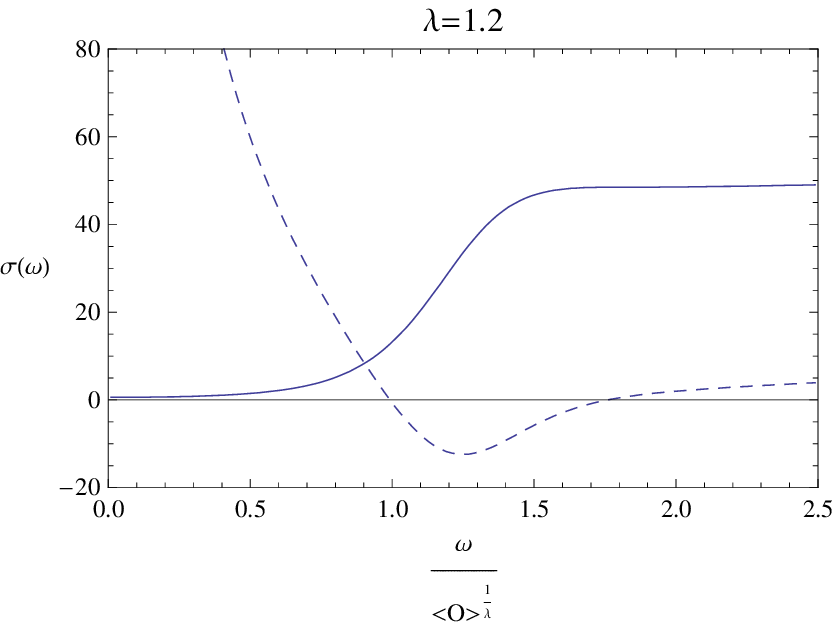}
\includegraphics[scale=0.3]{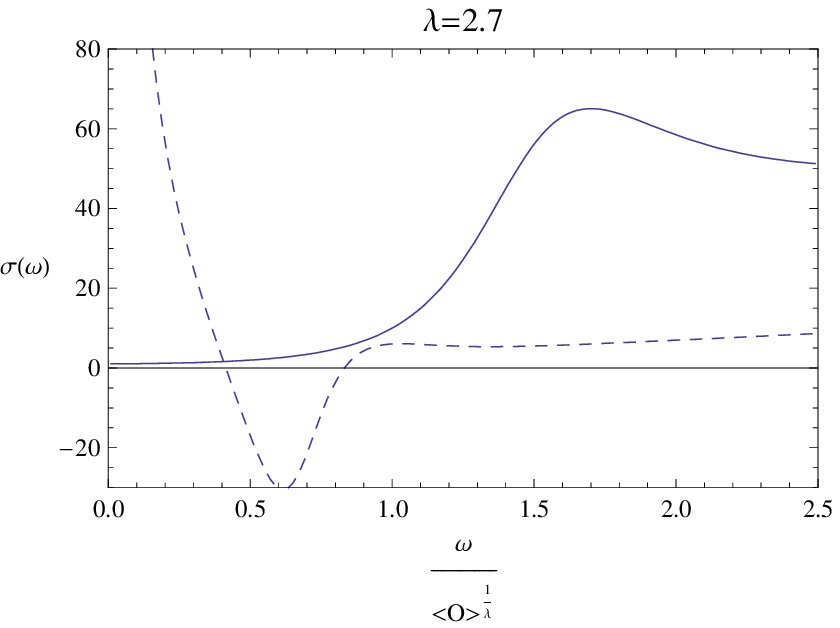}
\caption{Plots of the  conductivity versus the  frequency normalized by the condensate in the $3+1$ boundary theory. Each of the plots was calculated at $T/T_{c}=0.3$  and they are  labelled by the dimension of the dual condensates.}
\label{Figure6}
\end{figure}

\providecommand{\href}[2]{#2}\begingroup\raggedright\endgroup
\end{document}